\documentclass{jpp}

\usepackage{graphicx}
\usepackage{natbib}
\usepackage{bm}
\usepackage{amsmath}
\ifCUPmtlplainloaded \else
  \checkfont{eurm10}
  \iffontfound
    \IfFileExists{upmath.sty}
      {\typeout{^^JFound AMS Euler Roman fonts on the system,
                   using the 'upmath' package.^^J}%
       \usepackage{upmath}}
      {\typeout{^^JFound AMS Euler Roman fonts on the system, but you
                   dont seem to have the}%
       \typeout{'upmath' package installed. JPP.cls can take advantage
                 of these fonts, if you use 'upmath' package.^^J}%
      }
  \else
  \fi
\fi

\ifCUPmtlplainloaded \else
  \checkfont{msam10}
  \iffontfound
    \IfFileExists{amssymb.sty}
      {\typeout{^^JFound AMS Symbol fonts on the system, using the
                'amssymb' package.^^J}%
       \usepackage{amssymb}%
         \let\leq=\leqslant
         \let\geq=\geqslant
      }{}
  \fi
\fi

\ifCUPmtlplainloaded \else
  \IfFileExists{amsbsy.sty}
    {\typeout{^^JFound the 'amsbsy' package on the system, using it.^^J}%
     \usepackage{amsbsy}}
    {\providecommand\boldsymbol[1]{\mbox{\boldmath $##1$}}}
\fi

\newcommand\hatR{\skew3\hat{R}}      

\newsavebox{\astrutbox}
\sbox{\astrutbox}{\rule[-5pt]{0pt}{20pt}}
\newcommand{\astrut}{\usebox{\astrutbox}}

\newcommand\shalf{\ensuremath{{\scriptstyle\frac{1}{2}}}}

\title[Linear stability of field-aligned incompressible equilibria with anisotropic pressure]{On the linear stability of anisotropic pressure equilibria with field-aligned incompressible flow}

\author[A. Evangelias and G. N. Throumoulopoulos]%
{A.\ns E\ls V\ls A\ls N\ls G\ls E\ls L\ls I\ls A\ls S$^1$ \break
\and G.\ns N.\ns T\ls H\ls R\ls O\ls U\ls M\ls O\ls U\ls L\ls O\ls P\ls O\ls U\ls L\ls O\ls S$^1$
}

\affiliation{$^1$Department of Physics, University of Ioannina, Section of Astrogeophysics, \\GR 451 10 Ioannina, Greece\\[\affilskip]
\vspace{3mm}
Emails: aevag@cc.uoi.gr, gthroum@uoi.gr
}

\pubyear{2010}
\volume{650}
\pagerange{119--126}
\date{xxxxxxxx}
\begin{document}

\maketitle

\begin{abstract}
\begin{center}
ABSTRACT
\end{center}
We derive a sufficient condition for the linear stability of plasma equilibria with  incompressible flow parallel to the magnetic field, $\bf B$,   constant mass density and anisotropic pressure  such that the  quantity $\sigma_d= \mu_0(P_\parallel -P_\perp)/B^2$,  where $P_\parallel$ ($P_\perp$) is the pressure tensor element parallel (perpendicular) to $\bf B$,  remains constant. This condition is applicable  to any steady state without geometrical restriction. The condition, generalising  the respective condition for MHD equilibria with isotropic pressure and constant density derived in \cite{throumtaso},    involves   physically interpretable terms related to the magnetic shear, the flow shear and the variation of total pressure perpendicular to the magnetic surfaces. On the basis of this condition we prove that if a given equilibrium  is linearly stable, then the ones resulting from the application of Bogoyavlenskij symmetry transformations are linearly stable too, provided that a parameter involved in those  transformations is positive. In addition, we examine the impact of pressure anisotropy,  flow, and  torsion of  a helical  magnetic axis, for a specific class of analytic equilibria. In this case we find that the  pressure anisotropy  and the flow may have either stabilising or destabilising effects. Also,  helical configurations with small torsion and large  pitch seem to have more favorable stability properties. 
\end{abstract}
\vspace{4mm}
\hrule

\section{Introduction}\label{1}
For favorable confinement the  equilibrium states of fusion plasmas  are  desirable to be stable and therefore their stability study is of great importance.  Such low entropy states are  susceptible to numerous instabilities as strong ideal pressure and current driven modes, resistive instabilities often associated with magnetic reconnection, and kinetic micro-instabilities which occur when the distribution functions depart  from Mawxellians. Also, certain instabilities  are a source of turbulence, which  can drive transport. Investigation of macro-instabilities is usually performed within the framework of the ideal magnetohydrodynamics (MHD), since this model is a good approximation in describing the plasma as a macroscopic fluid and capturing most of the  physics of the force balance.
 
 There are two main methods for studying  MHD stability: first,  for small perturbations from the equilibrium,  linear stability is examined by the normal mode analysis which can calculate the perturbation growth rate and second,  the use of variational principles, involving perturbations of arbitrary amplitude and therefore covering the nonlinear regime,  in connection with  the sign of the  perturbation potential energy.   For static equilibria \cite{Bernstein} have derived a well known  Energy Principle which provides necessary and sufficient conditions for linear stability.

 It is also known that  plasma flows in magnetic confinement devices,  either induced  by external heating sources such as electromagnetic waves and neutral beams or developed intrinsically,  play an important role in the  formation of transport barriers and the transitions to improved confinement regimes, as the L-H transition.  In the presence of  flow, however, the problem of stability becomes much tougher because of the  antihermician convective flow term in the momentum equation.  As a result, only sufficient conditions for the linear stability of stationary equilibria were previously obtained: \cite{frieman,vlad1,vlad2,hameiri,throumtaso}. Particularly, in connection with the present study, the derivation  of   a sufficient condition for the linear stability of ideal MHD equilibria and plasmas of  constant  density, isotropic pressure and incompressible flow parallel to the  magnetic field, was initiated  in \cite{vlad1}, \cite{vlad2} and completed in \cite{throumtaso}. A key element to obtain this  condition is that the  pressure perturbation remains arbitrary, that is, there is no need to express that perturbation in terms of the Lagrangian displacement vector.

 Also, in  high temperature plasmas the collision time is so long that collisions can be neglected. Owing to the low collision frequency, a high temperature  plasma remains for long anisotropic, once anisotropy is induced, e.g.  by external heating sources. Macroscopic equations for a collisionless plasma with pressure anisotropy have been derived in \cite{CGL}. A detailed review of the anisotropic Chew-Goldberger-Low (CGL) model as well as of other collisionless fluid models is provided in \cite{hunana}. Pressure anisotropy is usually responsible for various  instabilities, such as the fire-hose and the mirror instability. Therefore, investigation of the stability properties of anisotropic plasmas, either static or stationary, is also a significant objective.

 In the present work we derive a sufficient condition for the linear stability of  equilibria with  field-aligned incompressible flows associated with plasmas of  constant density and constant  anisotropy function, $\sigma_d$,  working along the same lines as
in \cite{throumtaso}. Then, we examine the stability properties of such kind of anisotropic equilibria.
Specifically, in \S\ref{2} we present the basic dynamical MHD equations with anisotropic pressure and their equilibrium counterparts, while in \S\ref{3} we linearise the time dependent equations for small perturbations around the equilibrium quantities, and obtain a pertinent functional of the perturbation potential energy. In \S\ref{4} the sufficient condition is obtained  in the presence of pressure anisotropy, which generalises the one of  \cite{throumtaso} being valid for isotropic pressure. To derive that condition  it is also important that the perturbation of the effective pressure, $\mathcal{P}=(P_\parallel+P_\perp)/2$, remains arbitrary. Furthermore, in \S\ref{5} we show that whenever a given   equilibrium is linearly stable, then all families of equilibria derived  by  the symmetry transformations introduced in \cite{bogo1,eva1}, are also linearly stable provided that  a  parameter involved in  those transformations is positive. In addition, in \S\ref{6} we apply the derived sufficient condition to a class of analytic helically symmetric equilibria in order  to examine the impact of pressure anisotropy,  flow, and certain geometrical parameters characterizing  helical configurations, such as the  torsion and pitch length, on the stability of those equilibria.  Finally, \S\ref{7} summarizes the main results.
 
\section{Basic equations and background equilibria}\label{2}
The dynamics of a perfectly conducting plasma with anisotropic pressure is determined by the following set of equations:
  \begin{eqnarray}
   \label{CGLgeneral}
   \varrho ^{*}(D_t{\bf{V}^{*}})={\bf{J}^{*}}\times{\bf{B}^{*}}-{\nabla}\cdot{{\mathsfbi{P}^{*}}}, \quad \partial _t \varrho ^{*}+\nabla \cdot (\varrho ^{*}{\bf{V}^{*}})=0, \nonumber \\
   {\nabla}\cdot{\bf{B}^{*}}=0, \quad {\nabla}\times{\bf{B}^{*}}=\mu_{0}{\bf{J}^{*}}, \quad \partial _t {\bf{B}^{*}}={\nabla}\times({\bf{V}^{*}}\times {\bf{B}^{*}}).
\end{eqnarray}
Here,  ${\bf{B}^{*}}({\bf{r}},t)$ is the magnetic field, ${\bf{J}^{*}}({\bf{r}},t)$  the current density, $\mu _0$  the magnetic constant, ${\bf{V}^{*}}({\bf{r}},t)$  the plasma velocity, $\varrho ^{*}({\bf{r}},t)$  the mass density, and ${\mathsfbi{P}^{*}}$ the  pressure tensor  defined as
\begin{equation}
\label{tensor1}
{\mathsfbi{ P}^{*}}:=P_{\perp}^{*}{\mathsfbi{ I}}+\frac{\sigma_{d}^{*}}{\mu_{0}}{\bf{B}^{*}}{\bf{B}^{*}},
\end{equation}
where $\mathsfbi{ I}$ is the unit tensor; $\partial _t$ denotes the time derivative, while the Langrangian derivative is defined as $D_t\equiv \partial _t+({\bf{V}^{*}}\cdot \nabla)$.
The pressure tensor is diagonal in a local rectangular system with respect to ${\bf{B}^{*}}$, and consists of the scalar pressure elements  $P_{\parallel}^{*}({\bf{r}},t)$  and $P_{\perp}^{*}({\bf{r}},t)$ along and across ${\bf{B}^{*}}({\bf{r}},t)$, respectively,  while the dimensionless function
\begin{equation}
\label{sigma1}
\sigma _d^{*}:=\frac{\mu _0(P_{\parallel}^{*}-P_{\perp}^{*})}{|{\bf{B}^{*}}|^2}
\end{equation}
is   a measure of the pressure anisotropy. Particle collisions equilibrating parallel and perpendicular energies  lower  the value of  $\sigma _d^{*}$, and therefore  a highly collisional plasma is described accurately by a single scalar pressure, $P_{isotropic}^{*}$. In view of this fact, when pressure anisotropy is present it is useful to introduce an effective isotropic pressure, 
\begin{equation}
\label{peff}
\mathcal{P}^{*}({\bf{r}},t):=\frac{P_{\parallel}^{*}+P_{\perp}^{*}}{2},
\end{equation}
that reduces to $P_{isotropic}^{*}$ in the absence of anisotropy. By substituting equations (\ref{sigma1}) and (\ref{peff}) into the momentum equation of  set (\ref{CGLgeneral}) we obtain
\begin{equation}
\varrho ^{*}(D_t{\bf{V}^{*}})=(1-\sigma _d ^{*}){\bf{J}^{*}}\times{\bf{B}^{*}}-{\nabla}{{\mathcal{P}^{*}}}-\frac{{\bf{B}^{*}}}{\mu _0}({\bf{B}^{*}}\cdot \nabla \sigma _d ^{*})+\frac{|{\bf{B}^{*}}|^2}{2\mu _0}\nabla \sigma _d ^{*},
\label{momentum0}
\end{equation}
in which the scalar pressures $P_{\parallel}^{*}$ and $P_{\perp}^{*}$ do not appear explicitly. This is very useful for the stability analysis to follow in the next sections.

 The counterpart to set (\ref{CGLgeneral}) equilibrium equations are:
 \begin{eqnarray}
   \label{CGLpar}
   \varrho({\bf{V}}\cdot{\nabla}){\bf{V}}={\bf{J}}\times{\bf{B}}-{\nabla}\cdot{{\mathsfbi{P}}}, \quad {\nabla}\cdot (\varrho{\bf{V}})=0, \nonumber \\
   {\nabla}\cdot{\bf{B}}=0, \quad {\nabla}\times{\bf{B}}=\mu_{0}{\bf{J}}, \quad {\nabla}\times({\bf{V}}\times {\bf{B}})=0,
\end{eqnarray}
where the absence of the superscript $*$ denotes equilibrium and therefore not dependence  on  time. 

In the special case of collinear velocity and magnetic fields  related through
\begin{equation}
\label{collinear}
{\bf{V}}=\frac{\lambda({\bf{r}})}{\sqrt{\mu _0 \varrho}}{\bf{B}},
\end{equation}
 with $\lambda$  an arbitrary dimensionless scalar function, the continuity equation of the set (\ref{CGLpar}) takes the form $({\bf{B}}\cdot \nabla \varrho)/\varrho=-2 ({\bf{B}}\cdot \nabla \lambda)/\lambda$; substitution of the latter relation in the counterpart to (\ref{momentum0}) equilibrium momentum equation yields
 \begin{equation}
 \label{momentum1}
 (1-\sigma _d-\lambda ^2){\bf{J}}\times{\bf{B}}=\nabla \left(\mathcal{P}+\lambda ^2\frac{{\bf{B}}^2}{2\mu _0}\right)+\frac{{\bf{B}}^2}{2\mu _0}\nabla(1-\sigma _d-\lambda ^2)-\frac{{\bf{B}}}{\mu _0}\left[{\bf{B}}\cdot \nabla(1-\sigma _d-\lambda ^2)\right].
 \end{equation}
  Equation (\ref{momentum1})  is valid for arbitrary functions $\varrho$ and $\sigma _d$, and therefore for compressible flows. 
 
 Furthermore, we assume that  the mass density and the anisotropy function remain constant everywhere inside a volume  $\mathcal{D}$: $\varrho, \, \sigma_d =\mbox{const.}$.  Consequently, the corresponding field-aligned equilibrium flow becomes  incompressible and the function $\lambda$ is constant on magnetic surfaces $\psi =\mbox{const.}$, whenever such surfaces exist. Also,   both the magnetic and velocity field lie on those surfaces, ${\bf{B}}\cdot \nabla \lambda(\psi)={\bf{V}}\cdot \nabla \lambda(\psi)=0$. It is known that the existence of three-dimensional equilibria with nested toroidal magnetic surfaces is not guaranteed, and in general, irrespective of the existence of magnetic surfaces the function $\lambda$ is constant on both the magnetic field lines and the velocity streamlines. Henceforth, we will presume the existence of well defined equilibrium magnetic surfaces.
 Under these assumptions equation (\ref{momentum1}) takes  the simpler form
 \begin{equation}
 \label{momentum2}
 (1-\sigma _d-\lambda ^2){\bf{J}}\times{\bf{B}}=\nabla \mathcal{P}_s-(\lambda ^2)^{'}\frac{{\bf{B}}^2}{2\mu _0}\nabla \psi,
 \end{equation}
 where  the prime denotes differentiation with respect to $\psi$ and $\mathcal{P}_s$ is the total effective pressure in the absence of  flow $(\lambda =0)$, defined as
 \begin{equation}
 \label{ps}
 \mathcal{P}_s:=\mathcal{P}+\lambda ^2 \frac{{\bf{B}}^2}{2\mu _0}.
 \end{equation}
 Moreover, projection of equation (\ref{momentum2}) along ${\bf{B}}$ implies that the static total effective pressure is a surface quantity, $\mathcal{P}_s=\mathcal{P}_s(\psi)$, and therefore, the momentum equation (\ref{momentum2}) can be cast into the useful form
  \begin{equation}
  \label{momentumf}
  {\bf{J}}\times{\bf{B}}=g(\psi, {\bf{B}}^2)\nabla \psi ,
  \end{equation}
  where
  \begin{equation}
  \label{gpsi}
  g(\psi, {\bf{B}}^2):=(1-\sigma _d-\lambda ^2)^{-1}\left[\mathcal{P}_s^{'}-(\lambda ^2)^{'}\frac{{\bf{B}}^2}{2\mu _0}\right].
  \end{equation}
  Equation (\ref{momentumf}) implies that for equilibria with  field-aligned incompressible flows, constant density and constant anisotropy function, the current surfaces coincide with the magnetic surfaces, that is, the vectors  ${\bf{J}}$, ${\bf{B}}$ lie on common surfaces $\psi=\mbox{const.}$. In the subsequent sections we examine the stability of the aforementioned equilibria to small three-dimensional perturbations. To this end, we define the following quantities:
  \begin{equation}
  \label{NM}
  {\bf{N}}:={\bf{J}}\times{\bf{B}}, \quad {\bf{M}}:=\nabla \times {\bf{N}}=\nabla g\times \nabla \psi,
  \end{equation}
  from which it follows that 
  \begin{equation}
  \label{MN1}
  {\bf{N}}\cdot {\bf{M}}=0.
  \end{equation}
\section{Energy principle and perturbation potential energy}\label{3}
In order to examine the linear stability of a given equilibrium with anisotropic pressure and incompressible flows, $\nabla \cdot {\bf{V}}=0$, we assume that the equilibrium position, ${\bf{r}}$, is perturbed to a position ${\bf{r}}^{*}({\bf{r}},t)$, through the usual Lagrangian displacement vector  ${\bm{\xi}}({\bf{r}},t)\equiv {\bf{r}}^{*}-{\bf{r}}$, so that
\begin{eqnarray}
   \label{perturbations}
   {\bf{B}}^{*}={\bf{B}}({\bf{r}})+{\bf{b}}({\bf{r}},t), \quad {\bf{V}}^{*}={\bf{V}}({\bf{r}})+{\bf{v}}({\bf{r}},t), \quad {\bf{J}}^{*}={\bf{J}}({\bf{r}})+{\bf{j}}({\bf{r}},t),\nonumber \\
   {\mathcal{P}}^{*}={\mathcal{P}}({\bf{r}})+p({\bf{r}},t), \quad {\varrho}^{*}={\varrho}({\bf{r}})+{\delta}({\bf{r}},t), \quad {\sigma _d}^{*}={\sigma _d}({\bf{r}})+{\epsilon}({\bf{r}},t).
\end{eqnarray}
Here,  ${\bf{b}}, \, {\bf{j}}, \, p, \, \delta, \,\epsilon$, and
\begin{equation}
\label{veloperp}
{\bf{v}}={\bf{u}}({\bf{r}},t)+\frac{\partial {\bm{\xi}}}{\partial t}
\end{equation}
correspond to small perturbations of the respective equilibrium quantities. Note that we have assumed perturbations of the effective pressure, ${\mathcal{P}}$, and the anisotropy function, $\sigma _d$, instead of explicit perturbations of the scalar pressures $P_{\parallel}$ and $P_{\perp}$. Also, on the fixed boundary  $\partial \mathcal{D}$ surrounding a plasma domain $\mathcal{D}$ of interest we adopt the  following  conditions:
\begin{equation}
\label{boundary}
{\bf{b}}\cdot {\bf{\hat{n}}}={\bf{u}}\cdot {\bf{\hat{n}}}=0,
\end{equation}
where  ${\bf{\hat{n}}}$ is the perpendicular outward unit vector on the boundary.
Introducing perturbations  (\ref{perturbations}) into the dynamical equations (\ref{CGLgeneral}) and employing the equilibrium equations  (\ref{CGLpar}),   we obtain the following linearised equations 
\begin{eqnarray}
   \label{linearised}
   \nabla \cdot {\bf{b}}=0, \quad {\bf{j}}=\frac{1}{\mu _0}\nabla \times {\bf{b}}, \quad \frac{\partial \delta}{\partial t}+{\bf{V}}\cdot \nabla \delta +\nabla \cdot (\varrho {\bf{v}}) = 0, \nonumber \\
   \varrho \frac{\partial {\bf{v}}}{\partial t}+\varrho \left[ ({\bf{V}}\cdot \nabla ){\bf{v}}+({\bf{v}}\cdot \nabla ){\bf{V}}\right]-(1-\sigma _d)({\bf{J}}\times {\bf{b}}+{\bf{j}}\times {\bf{B}})+\nabla p = G({\bf{r}},t),
\end{eqnarray}
where
 
 \begin{eqnarray}
G & := & -\delta ({\bf{V}}\cdot \nabla ){\bf{V}}-\epsilon {\bf{J}}\times {\bf{B}}+\frac{1}{\mu _0}\left[ \frac{{\bf{B}}^2}{2}\nabla \epsilon -({\bf{B}}\cdot \nabla \epsilon +{\bf{b}}\cdot \nabla \sigma _d){\bf{B}}
  \right. \nonumber\\
&& \left. +\left({\bf{B}}\cdot {\bf{b}}+\frac{{\bf{b}}^2}{2}\right)\nabla\sigma _d-({\bf{B}}\cdot \nabla \epsilon){\bf{b}}\right].
\label{G}
\end{eqnarray}
 
 At this point we note that in order that the  set of dynamical equations (\ref{CGLgeneral}) be closed  one needs in connection with the  pressures $P_\parallel^{*}$ and $P_\perp^{*}$  a couple of energy equations or equations of state, e.g. the double adiabatic equations \cite{CGL} associated with the known CGL model.  In the present study  one such equation of state corresponds to  incompressibility in connection with an evolution with  constant mass density, $\varrho^{*}=\varrho=\mbox{const.}, \ (\delta=0)$. Also, the fact that the momentum equation can be cast in the form (\ref{momentum0}) involving the pressures  $P_\parallel^{*}$ and $P_\perp^{*}$ only implicitly through the effective pressure, $\mathcal{P}^{*}$, and the anisotropy function, $\sigma_d$, motivated us to adopt as second equation of state the constrain that the latter  function remains constant,  $ \sigma _d^{*}=\sigma _d=\mbox{const.}, \ (\epsilon =0)$. This implies that  $P_\parallel^{*}$ and $P_\perp^{*}$ (precisely their difference)  evolve in such a way that they keep proportional to the magnetic pressure $B^2/(2\mu_0)$, which on physical grounds is an acceptable approximation. Consequently,  equation (\ref{G}) implies that $G=0$, while the second of the linearised equations  (\ref{linearised}) reduces to
\begin{equation}
\label{incop1}
\nabla \cdot {\bf{v}}=0.
\end{equation}
In view of this relation, we consider incompressible perturbations, $\nabla \cdot {\bm{\xi}}=0$, such that the condition ${\bm{\xi}}\cdot {\bf{\hat{n}}}=0$ is satisfied on the boundary.   Then, equation (\ref{incop1})  implies that $\nabla \cdot {\bf{u}}=0$.
Subsequently, the perturbations of the magnetic and  velocity fields can be expressed in terms of ${\bm{\xi}}$, as
\begin{equation}
\label{bu}
{\bf{b}}=\nabla \times ({\bm{\xi}}\times {\bf{B}}), \quad {\bf{u}}=\nabla \times ({\bm{\xi}}\times {\bf{V}}),
\end{equation}
while the linearised force balance equation of the set (\ref{linearised}) is put in the form
\begin{equation}
\label{forcebalance}
\varrho \frac{\partial ^2 {\bm{\xi}}}{\partial t^2}+2\varrho({\bf{V}}\cdot \nabla)\frac{\partial {\bm{\xi}}}{\partial t}+\nabla f={\bf{\hat{F}}}{\bm{\xi}}.
\end{equation}
Here,
\begin{equation}
\label{operator}
{\bf{\hat{F}}}:=\varrho {\bf{u}}\times (\nabla \times {\bf{V}}) +\varrho {\bf{V}}\times (\nabla \times {\bf{u}})+(1-\sigma _d){\bf{J}}\times {\bf{b}}-\frac{1}{\mu _0}{\bf{B}}\times (\nabla \times {\bf{b}}),
\end{equation}
 is the symmetric force operator and the scalar function $f$ is defined as
\begin{equation}
\label{f}
f:= \varrho {\bf{V}}\cdot {\bf{u}}+p.
\end{equation}

The energy principle is based of the fact that the total perturbation energy 
\begin{equation}
\label{E}
E=K+W=\frac{1}{2} \int _\mathcal{D} \varrho  \dot{\bm{\xi}}^2d^3 r - \frac{1}{2} \int _\mathcal{D} \bm{\xi}\cdot {\bf \hat{F}}
\bm{\xi}d^3 r 
\end{equation}
is conserved, where  $K$ is the kinetic energy and  $W$ the potential energy.  Stability is related to the sign of $E$. Since, $K$ is quadratic in velocity,  and therefore non negative definite,  a sufficient condition for a given equilibrium to be linearly stable is $W\geq 0$.  
On account of equations (\ref{boundary})-(\ref{f}), $W$ is expected to depend only on the physical quantities of the background equilibrium and the displacement vector, $\bm{\xi}$,  with the exception of the perturbation, $p$, of the effective pressure, appearing in the gradient of the quantity $f$. However, the contribution of the latter term to $W$ vanishes  due to the  implied boundary conditions on $\partial \mathcal{D}$.  Thus, one finds for the perturbation potential energy
\begin{equation}
\label{pote1}
W=\frac{1}{2}\int _\mathcal{D}\left\lbrace \varrho {\bf{u}}\cdot [{\bm{\xi}}\times (\nabla \times {\bf{V}})]-\varrho {\bf{u}}^2+(1-\sigma _d){\bf{b}}\cdot({\bf{J}}\times {\bm{\xi}})+\frac{(1-\sigma _d)}{\mu _0}{\bf{b}}^2 \right\rbrace d^3 {\bf{r}}.
\end{equation}
Now we focus our study on the investigation of the linear stability for the  equilibria with field-aligned incompressible flows, defined by  equation (\ref{collinear}). In this case  (\ref{pote1}) becomes
\begin{equation}
\label{pote2}
W=\frac{1}{2\mu _0}\int _\mathcal{D}\left\lbrace (1-\sigma _d-\lambda ^2) [{\bf{b}}^2+ {\bf{b}}\cdot (\mu _0 {\bf{J}}\times {\bm{\xi}})]-2\lambda ({\bm{\xi}}\cdot \nabla \lambda)({\bm{\xi}}\cdot [({\bf{B}}\cdot \nabla ){\bf{B}}]) \right\rbrace d^3 {\bf{r}}.
\end{equation}
In the next section we employ the form of $W$ given in equation (\ref{pote2}) to derive a sufficient condition for the linear stability of the respective kind of equilibria.

\section{Sufficient condition for linear stability}\label{4}

It is recalled  that for the equilibria with field-aligned incompressible flows, constant mass density and constant pressure anisotropy function, $\sigma_d$,  defined by equations (\ref{collinear})-(\ref{NM}), the current density stays on  magnetic surfaces, $\psi=\mbox{const.}$, and thus, the vectors ${\bf{B}}$, ${\bf{J}}$ and ${\bf{N}}={\bf{B}}\times {\bf{J}}$, form a basis in $\mathbb{R}^3$ space. Accordingly, following the analysis in \cite{throumtaso}  we expand the displacement vector in this basis, as
\begin{equation}
\label{basis}
{\bm{\xi}}=\alpha({\bf{r}},t){\bf{N}}+\beta({\bf{r}},t){\bf{J}}+\gamma({\bf{r}},t){\bf{B}},
\end{equation}
where $\alpha$, $\beta$, and $\gamma$ are arbitrary, appropriately dimensional scalar functions. We note that in \cite{throumtaso}   a sufficient condition  was derived for the linear stability of equilibria with field-aligned incompressible flows, isotropic pressure  and  constant density; in fact, the constant density and the vacuum  magnetic permeability constant were set to unity therein. These  equilibria  are recovered   from  the respective anisotropic-pressure equilibria described in \S\ref{2}  for $\sigma _d=0$. Also, we observe that the form  (\ref{pote2}) of potential energy for $\sigma _d=0$ (and $\varrho=\mu _0=1$)  reduces to the respective form \citep[equation (13)]{throumtaso}. Thus, it is straightforward to derive along the same lines as in \cite{throumtaso} a sufficient condition for the linear stability of the present anisotropic equilibria;  specifically,  we 
decompose  the potential energy of equation (\ref{pote2}) into two integrals as
\begin{equation}
\label{two}
W=W_1+W_2,
\end{equation}
and following step by step the procedure in \citep[Appendix]{throumtaso}, we obtain 
\begin{equation}
\label{W1}
W_1=\frac{1}{2\mu _0}\int _\mathcal{D}(1-\sigma _d-\lambda ^2)({\bf{b}}+\alpha \mu _0 {\bf{J}}\times {\bf{N}})^2 d^3{\bf{r}},
\end{equation}
and
\begin{equation}
\label{W2}
W_2=\frac{1}{2\mu _0}\int _\mathcal{D}\mathcal{A}(\sqrt{2}g\alpha)^2d^3{\bf{r}},
\end{equation}
where
\begin{eqnarray}
\label{Alpha}
\mathcal{A} & := & -(1-\sigma _d-\lambda ^2)\left\{ \astrut |\mu _0 {\bf{J}}\times \nabla \psi |^2-(\mu _0 {\bf{J}}\times \nabla \psi)\cdot [(\nabla \psi \cdot \nabla){\bf{B}}]\astrut \right\}\nonumber \\
&& +\frac{\mu _0}{2}[\ln(1-\sigma _d-\lambda ^2)]^{'}|\nabla \psi |^2\nabla \psi \cdot \nabla \left(P_{\perp}+\frac{{\bf{B}}^2}{2 \mu _0}\right).
\end{eqnarray}
Equations (\ref{W1}) and (\ref{W2}) imply that $W$ is non-negative if both quantities $1-\sigma _d-\lambda ^2$ and $\mathcal{A}$ are also non-negative definite in $\mathcal{D}$. Thus, we conclude to the following statement: 
\textit{An  equilibrium with anisotropic pressure with  $\sigma_d=\mbox{const.}$, field-aligned incompressible flow in connection with constant plasma density is linearly stable  if both of the following conditions are satisfied: }
\begin{equation}
\label{con1}
1-\sigma _d -\lambda ^2 \geq 0,
\end{equation}
\begin{equation}
\label{con2}
\mathcal{A} \geq 0.
\end{equation}
Conditions (\ref{con1}) and (\ref{con2})  can be applied to any steady state without geometrical restriction. 
They  generalise  the ones derived  in \cite{throumtaso} for isotropic pressure, since for $\sigma _d=0$ condition (\ref{con1}) reduces to sub-Alfv\' enic flows, $\lambda ^2<1$, while  (\ref{con2}) reduces to condition \citep[equation (22)]{throumtaso}. In fact,  the expression (\ref{Alpha})  is more compact because it consists of three terms, instead of four  terms in  \citep[equation (22)]{throumtaso}.  The first of these terms,
\begin{equation}
\label{A1}
A_1= -(1-\sigma _d-\lambda ^2)|\mu _0 {\bf{J}}\times \nabla \psi |^2, 
\end{equation}
is negative and therefore  always destabilising, in potential connection with current-driven modes. The second term, 
\begin{equation}
\label{A2}
A_2=(1-\sigma _d-\lambda ^2)(\mu _0 {\bf{J}}\times \nabla \psi)\cdot [(\nabla \psi \cdot \nabla){\bf{B}}],  
\end{equation}
is related to the magnetic shear (i.e. it depends on the variation of ${\bf{B}}$ across the magnetic surfaces), and can be either stabilising or destabilising. The third term,
\begin{equation}
\label{A3}
A_3=\frac{\mu _0}{2}[\ln(1-\sigma _d-\lambda ^2)]^{'}|\nabla \psi |^2\nabla \psi \cdot \nabla \left(P_{\perp}+\frac{{\bf{B}}^2}{2 \mu _0}\right),
\end{equation}
 can be regarded  as flow term, though being affected by anisotropy, since it vanishes in the absence of flow   ($\lambda =0$). Note that   it relates to the variation of the total  pressure perpendicular  to the magnetic surfaces; indeed,  on account of equations (\ref{collinear}) and (\ref{ps}) one finds  
\begin{equation}
\label{pressuremeans}
P_{\perp}+\frac{{\bf{B}}^2}{2 \mu _0}=\underbrace{\mathcal{P}_s(\psi)}_\text{static}-\underbrace{\frac{1}{2}\varrho{\bf{V}}^2}_\text{flow}+\underbrace{(1-\sigma _d)\frac{{\bf{B}}^2}{2 \mu _0}}_\text{magnetic},
\end{equation}
which involves all three pressures, static effective, flow and magnetic,  the latter  being influenced  by the pressure anisotropy through the factor ($1-\sigma_d$).  In addition, satisfaction of condition (\ref{con1}) in the absence of flow,  $\sigma_d\leq 1$, implies  that the corresponding static anisotropic equilibria  are  stable under the fire-hose instability \citep[see][]{cldo}.

Before closing this section we note  that every equilibrium state with incompressible flow  and anisotropy function uniform  on the magnetic surfaces, $\sigma_d=\sigma_d(\psi)$ having  certain continuous geometrical symmetry is governed by a generalised Grad-Shafranov (GS) equation for the flux function $\psi$, e.g. equation (28) in \cite{eva2} governing helically symmetric equilibria. That  equation contains a quadratic  $|\nabla \psi|^2$-term. For this reason we have introduced the integral transformation,
\begin{equation}
\label{Utrans}
U(\psi)=\int _0^{\psi} \sqrt{1-\sigma _d(x)-\lambda ^2(x)}dx,\quad  1-\sigma_d-\lambda^2\geq 0,
\end{equation}
under which the respective transformed GS equation, becomes  free of a quadratic term as $|\nabla U|^2$, and can be solved by analytical techniques in the $U$-space. Trasformation (\ref{Utrans}) does not change the topology of the magnetic surfaces, but just relabels them by the flux function $U$, and consists of a generalisation of the transformation in  \cite{clemente} for static anisotropic equilibria $(\lambda=0)$  and that in  \cite{simintzis} for respective stationary, isotropic equilibria $(\sigma _d=0)$. Under transformation (\ref{Utrans}),  condition (\ref{con1}) is trivially satisfied, while condition (\ref{con2}), valid for $\sigma _d=\mbox{const.}$,  is expressed in $U$-space as
\begin{eqnarray}
\label{Au}
\mathcal{A}& = &-|\mu _0 {\bf{J}}\times \nabla U |^2+(\mu _0 {\bf{J}}\times \nabla U)\cdot [(\nabla U \cdot \nabla){\bf{B}}]\nonumber \\
&&+\frac{\mu _0}{2(1-\sigma _d-\lambda ^2)}\frac{d\ln(1-\sigma _d-\lambda ^2)}{dU}|\nabla U |^2\nabla U \cdot \nabla \left(P_{\perp}+\frac{{\bf{B}}^2}{2 \mu _0}\right).
\end{eqnarray}
Condition  $\mathcal{A}\geq 0$ in connection with  form (\ref{Au})  
will be employed  in \S\ref{6} to examine the stability of specific helically symmetric equilibria.

\section{Stability under symmetry transformations}\label{5}

In a recent work, \cite{eva1}, a set of symmetry transformations that map  anisotropic plasma equilibria with field-aligned incompressible flows were introduced; specifically, these transformations when applied to a given equilibrium with parallel incompressible flow and anisotropy function, $\sigma_d$, uniform on the magnetic surfaces labeled by the function 
$\psi$,  $\{ {\bf{B}}, \, \varrho(\psi) , \, {\bf{V}}=\lambda{\bf{B}}/(\sqrt{\mu_0 \varrho}), \, \mathcal{P}, \, \sigma_d(\psi) \}$, produce an infinite family  of respective equilibria with field-aligned incompressible flows, but density and anisotropy function that may vary on the magnetic surfaces, $\{ {\bf{B}_1}, \, \varrho _1, \, {\bf{V}}_1, \, \mathcal{P}_1, \, \sigma _{d_1} \}$.  These transformations  are defined by
\begin{eqnarray}
 \label{transfCGLpar}
  {\bf{B}}_1=\frac{b_1({\bf{r}})}{n_1({\bf{r}})}{\bf{B}},\quad {\bf{V}}_1=\frac{c_1({\bf{r}})\sqrt{1-\sigma _d}}{a_1({\bf{r}})\sqrt{\mu _0 \varrho}}{\bf{B}},\nonumber \\
 \varrho _1 ({\bf{r}})=a_1^2({\bf{r}})\varrho ,\quad \mathcal{P}_1=C\mathcal{P}+(1-\sigma _d)(C-b_1^2({\bf{r}}))\frac{{\bf{B}}^2}{2\mu _0},\nonumber \\
 \sigma _{d_1}=1-n_1^2({\bf{r}})(1-\sigma _d), \quad C= \frac{\left[b_1^2({\bf{r}})-c_1^2({\bf{r}})\right](1-\sigma _d)}{1-\sigma _d-\lambda ^2}=\mbox{const.}\neq 0,
 \end{eqnarray}
where $a_1, \, b_1, \, c_1$, and $n_1$ are  scalar functions, which must satisfy the relations
\begin{equation}
\label{valid}
{\bf{B}}\cdot \nabla \left(\frac{b_1}{n_1}\right)=0, \quad {\bf{B}}\cdot \nabla (a_1c_1)=0.
\end{equation}
Transformations (\ref{transfCGLpar}),  consisting a generalisation of those  introduced in \cite{bogo1} for respective equilibria with isotropic pressure,  preserve the topology of the magnetic surfaces,  $\psi \equiv \psi_1 = \mbox{const.}$,  between the original and the transformed equilibrium, i.e.  ${\bf{B}}\cdot \nabla \psi={\bf{B}}_1\cdot \nabla \psi$. It  was also proved in  \cite{eva1}  that these transformations can break the geometrical symmetry of a given equilibrium only when  the fields  ${\bf{B}}$ and ${\bf{V}}$ are purely poloidal.

From (\ref{transfCGLpar}) it readily follows  that the transformed collinear velocity and magnetic fields are related through
\begin{equation}
\label{collineartrans}
{\bf{V}}_1=\frac{\lambda _1}{\sqrt{\mu _0 \varrho _1}}{\bf{B}}_1, \quad \lambda _1=\frac{c_1 n_1}{b_1}\sqrt{1-\sigma _d},
\end{equation}
and thus, the trasformed equilibria satisfy a force-balance equation analogous to (\ref{momentum1}):
 \begin{equation}
 \label{momentum1transf}
 (1-\sigma _{d_1}-\lambda_1 ^2){\bf{J}}_1\times{\bf{B}}_1=\nabla \left(\mathcal{P}_1+\lambda _1^2\frac{{\bf{B}}_1^2}{2\mu _0}\right)+\frac{{\bf{B}}_1^2}{2\mu _0}\nabla(1-\sigma _{d_1}-\lambda_1 ^2)-\frac{{\bf{B}}_1}{\mu _0}\left[{\bf{B}}_1\cdot \nabla(1-\sigma _{d_1}-\lambda_1 ^2)\right].
 \end{equation}
 With the aid of equations (\ref{transfCGLpar}) and (\ref{collineartrans}) we obtain the useful relations
 \begin{equation}
 \label{s1l1}
 1-\sigma _{d_1}-\lambda _1^2=C\left(\frac{n_1}{b_1}\right)^2(1-\sigma _d-\lambda^2),
 \end{equation}
 \begin{equation}
 \label{p1l1}
 \mathcal{P}_1+\lambda _1^2\frac{{\bf{B}}_1^2}{2\mu _0}=C\left(\mathcal{P}+\lambda ^2\frac{{\bf{B}}^2}{2\mu _0} \right)=C\mathcal{P}_s,
 \end{equation}
 under which equation (\ref{momentum1transf}) reduces to 
 \begin{equation}
 \label{momentumtransf1}
\left(\frac{n_1}{b_1}\right)^2 {\bf{J}}_1\times{\bf{B}}_1={\bf{J}}\times{\bf{B}}+\frac{{\bf{B}}_1^2}{2\mu _0}\nabla \left(\frac{n_1}{b_1}\right)^2.
 \end{equation}
 Now,  presume that the original equilibrium belongs to the family described  in \S\ref{2} for which equation (\ref{momentumf}) holds; then equation (\ref{momentumtransf1}) becomes  
 \begin{equation}
 \label{momentumtransf2}
\left(\frac{n_1}{b_1}\right)^2 {\bf{J}}_1\times{\bf{B}}_1=g(\psi ,{\bf{B}}^2)\nabla \psi+\frac{{\bf{B}}_1^2}{2\mu _0}\nabla \left(\frac{n_1}{b_1}\right)^2.
 \end{equation}
 Projection of equation (\ref{momentumtransf2}) along ${\bf{J}}_1$ yields
 \begin{equation}
 \label{J1psi}
 {\bf{J}}_1\cdot \nabla \psi =-\frac{{\bf{B}}_1^2}{2\mu _0}\left[{\bf{J}}_1\cdot \nabla \left(\frac{n_1}{b_1}\right)^2\right],
 \end{equation}
 and  thus,  it turns out  that the transformed current density, ${\bf{J}}_1$, remains on the magnetic surfaces if and only if the ratio $n_1/b_1$ is uniform on those  surfaces
 \begin{equation}
 \label{constraint}
 \frac{n_1}{b_1}:= y(\psi).
 \end{equation}
 In this case, the transformed vectors, ${\bf{J}}_1, \, {\bf{B}}_1$, and ${\bf{N}}_1\equiv {\bf{J}}_1\times{\bf{B}}_1$, form a basis in $\mathbb{R}^3$, and thus, for constant  $\varrho _1$ and $\sigma _{d_1}$,  sufficient conditions for the linear stability of the transformed equilibria, satisfying  (\ref{transfCGLpar}), (\ref{collineartrans}) and (\ref{constraint}),  analogous to (\ref{con1}) and  (\ref{con2}) can be derived:
 \begin{equation}
\label{con1f}
1-\sigma _{d_1} -\lambda_1 ^2 \geq 0,
\end{equation}
\begin{equation}
\label{con2f}
\mathcal{A}_1 \geq 0,
\end{equation}
 where
 \begin{eqnarray}
\label{A}
\mathcal{A}_1&=&-(1-\sigma _{d_1}-\lambda _1 ^2)\{|\mu _0 {\bf{J}}_1\times \nabla \psi |^2+(\mu _0 {\bf{J}}_1\times \nabla \psi)\cdot [(\nabla \psi \cdot \nabla){\bf{B}}_1]\}\nonumber \\
&&+\frac{\mu _0}{2}[\ln(1-\sigma _{d_1}-\lambda _1^2)]^{'}|\nabla \psi |^2\nabla \psi \cdot \nabla \left(\mathcal{P}_{\perp _1}+\frac{{\bf{B}}_1^2}{2 \mu _0}\right).
\end{eqnarray}

 In order to investigate the stability of an aforementioned transformed equilibria, we presume  that the original equilibrium has constant density and anisotropy functions ($\varrho=\mbox{const.}$,  $\sigma _d=\mbox{const.}$) and is stable under small three-dimensional perturbations;  therefore,  conditions (\ref{con1}) and (\ref{con2}) are satisfied in $\mathcal{D}$. Note that in this case the scalar functions of  transformation (\ref{transfCGLpar}) must have a structrure of the form: $a_1=\mbox{const.}, \, n_1= \mbox{const.}, \, b_1=b_1(\psi), \, c_1=c_1(\psi)$;  as a result,  breaking  potential geometrical symmetry of the original equilibrium is not possible by the transformation even for purely poloidal ${\bf{B}}$ and  ${\bf{V}}$ fields. In \cite{vlad2} the stability of respective isotropic equilibria was examined; in particular, it was stated therein that all equilibrium families resulting from the application of the respective isotropic transformations introduced by  Bogoyavlenskij to given equilibria of field-aligned incompressible flows, isotropic pressure and constant mass density, are linearly stable, if either the original equilibria is stable and $C>0$, or the original equilibria is unstable and $C<0$. A straightforward calculation shows that 
 \begin{equation}
 \label{AA1}
 \mathcal{A}_1=C\mathcal{A},
 \end{equation}
so that conditions (\ref{con1f}) and (\ref{con2f}) assume the form
 \begin{equation}
\label{con1ff}
C\left( \frac{n_1}{b_1}\right)^2(1-\sigma _d -\lambda ^2) \geq 0,
\end{equation}
\begin{equation}
\label{con2ff}
C\mathcal{A} \geq 0.
\end{equation}
 By inspection of the later relations we come to the conclusion that the transformed equilibrium is linearly stable if, either (i) the original one is linearly stable and $C>0$, or (ii) neither of the conditions (\ref{con1}), (\ref{con2}) are satisfied and $C<0$. However, when $C$ is negative equation (\ref{p1l1}) in the absence of flow $(\lambda=0)$ yields  the  physically   unacceptable relation
 \begin{equation}
 \label{pps}
\frac{\mathcal{P}_1}{\mathcal{P}}<0.
 \end{equation}
Thus, we finally conclude to the formulation of the following statement: \\
  \textit{The infinite class of equilibria,  obtained from the application of the symmetry transformations (\ref{transfCGLpar}) for the case $\varrho _1=\mbox{const.}$ and $\sigma _{d_1}=\mbox{const.}$ to given respective equilibria which are linearly stable (by satisfying  the sufficient conditions (\ref{con1})-(\ref{con2})), are  also linearly stable if the transformation constant $C$ is positive definite.}
\newline
  This statement corrects and generalises the  respective statement  in \cite{vlad2} for isotropic equilibria $(\sigma _d=\sigma _{d_1}=0, \, n_1=1)$. 

\section{Linear stability of helically symmetric equilibria}\label{6}
In this section we consider helically symmetric equilibria with field-aligned incompressible flows and anisotropic pressure derived in \cite{eva2}, for which the following relations hold
\begin{eqnarray}
\label{helical}
{\bf{B}}=I(U){\bf{h}}+(1-\sigma _d-M_p(U)^2)^{-1/2}{\bf{h}}\times \nabla U(r,u), \quad {\bf{V}}=\frac{M_p(U)}{\mu _0 \varrho}{\bf{B}},\nonumber \\
\varrho =\mbox{const.}, \quad \sigma _d=\mbox{const.}, \quad \mathcal{P}=\mathcal{P}_s(U)-M_p^2(U)\frac{{\bf{B}}^2}{2\mu _0}.
\end{eqnarray}
Here the function $U(r,u)$ labels the magnetic surfaces, where $(r,u,z)$ are helical coordinates defined through the usual cylindrical ones $(\rho,\phi,\zeta)$ as $r=\rho$, $u=z-\eta\phi $, $z=\zeta$; $I$ relates to the helicoidal magnetic field; $\mathcal{P}_s$ is the static effective pressure;   $M_p$ is the Alfv\' en Mach function for the poloidal field, defined as $M_p:=(\sqrt{\mu _0 \varrho)}|{\bf{V}}_{pol}|)/{\bf{B}}_{pol}$,  which for parallel flows equals to the total Mach function ($M=\lambda=\sqrt{\mu _0 \varrho}|{\bf{V}}|/{\bf{B}})$.  All these functions are uniform on the magnetic surfaces. In addition, the vector ${\bf{h}}:=(-\eta/(k^2r^2+\eta^2)){\bf{g}}_{z}$ points along the helical direction, where $\eta$  is an arbitrary constant, and  ${\bf{g}}_{i}=\partial {\bf{r}}/\partial i \   (i=r,u,z)$ are  the covariant helical basis vectors. According to \cite{eva2}, equilibria (\ref{helical}) are governed by the following  generalised GS equation
\begin{equation}
\label{GS}
\mathcal{L}U+\frac{1}{2}\frac{d}{dU}[(1-\sigma _d-M_p^2)I^2]+\mu _0(k^2r^2+\eta^2)\frac{d\mathcal{P}_s}{dU}+\frac{2\eta}{k^2r^2+\eta^2}(1-\sigma _d-M_p^2)^{1/2}I=0,
\end{equation}
where the elliptic operator involved is defined as $\mathcal{L}:=(r^2+\eta^2)[\nabla \cdot(\nabla/(r^2+\eta^2)]$. In fact, equation (\ref{GS})  constitutes a reduced form of the more general one  \citep[equation (30)]{eva2}  valid in general for non-collinear   ${\bf{V}}$  and  ${\bf{B}}$, with the replacements of the parameters $m$ and $k$ therein by  $-\eta$ and $-1$, respectively, here. Under the following linearising ansatz for the free function terms, 
\begin{equation}
\label{ans}
(1-\sigma _d-M_p^2)^{1/2}I=w_1 U, \quad \mathcal{P}_s=w_2-2 w_3^2\frac{U^2}{\mu _0},
\end{equation}
the GS equation (\ref{GS}) reduces to the form
\begin{equation}
\label{lin}
\frac{1}{r^2}\frac{\partial ^2 U}{\partial u^2}+\frac{1}{r}\frac{\partial}{\partial r}\left( \frac{r}{r^2+\eta^2}\frac{\partial U}{\partial r}\right)+\left(\frac{w_1^2}{r^2+\eta^2}+\frac{2w_1\eta}{(r^2+\eta^2)^2}-4w_3^2\right)U=0,
\end{equation}
where $w_1, \, w_2$, and $w_3$ are arbitrary constants. PDE (\ref{lin}) was solved analytically in \cite{bogo2} and the exact solution obtained therein is
\begin{equation}
\label{bogsol}
U_{Nql}=e^{-w_3 r^2}\{f_N R_{0N}(s)+r^qR_{ql}(s)[c_{ql}\cos(qu/\eta)+d_{ql}\sin(qu/\eta)]\},
\end{equation}
where $N, \, q, \, l$ are arbitrary integers $\geq 0$ satisfying the condition $2N>2l+q$;  $f_N, \, c_{ql}, \, d_{ql}$ are arbitrary coefficients, and $s=2w_3r^2$; the form of the polynomial functions $R_{ql}(s)$ involve derivatives of the Laguerre polynomials $L_{q+l}(s)$ \citep[see][equations (3.7)-(3.8)]{bogo2}, and $R_{0N}$ is the respective polynomial for $q=0$, $l=N\neq 0$. On account of  solution $(\ref{bogsol})$,  different classes of exact helically symmetric MHD equilibria describing astrophysical jets with isotropic pressure $(\sigma _d=0)$ were constructed in \cite{bogo2}. For such kind of equilibria both the magnetic field, the flow velocity and the current density fall rapidly to zero at $r\rightarrow \infty$, while the pertinent isotropic pressure takes a limiting constant value therein.
We have to note that solution (\ref{bogsol}) can also accurately describe helically symmetric CGL anisotropic pressure equilibria, with $\sigma _d$ being a surface quantity (or constant), and incompressible flow. This is due to the fact that although the MHD and CGL models are established through different physical assumptions for the particle collisions, the respective  generalised GS equations governing them are identical in form.

\begin{figure}
 \centering
 \includegraphics[scale=0.75]{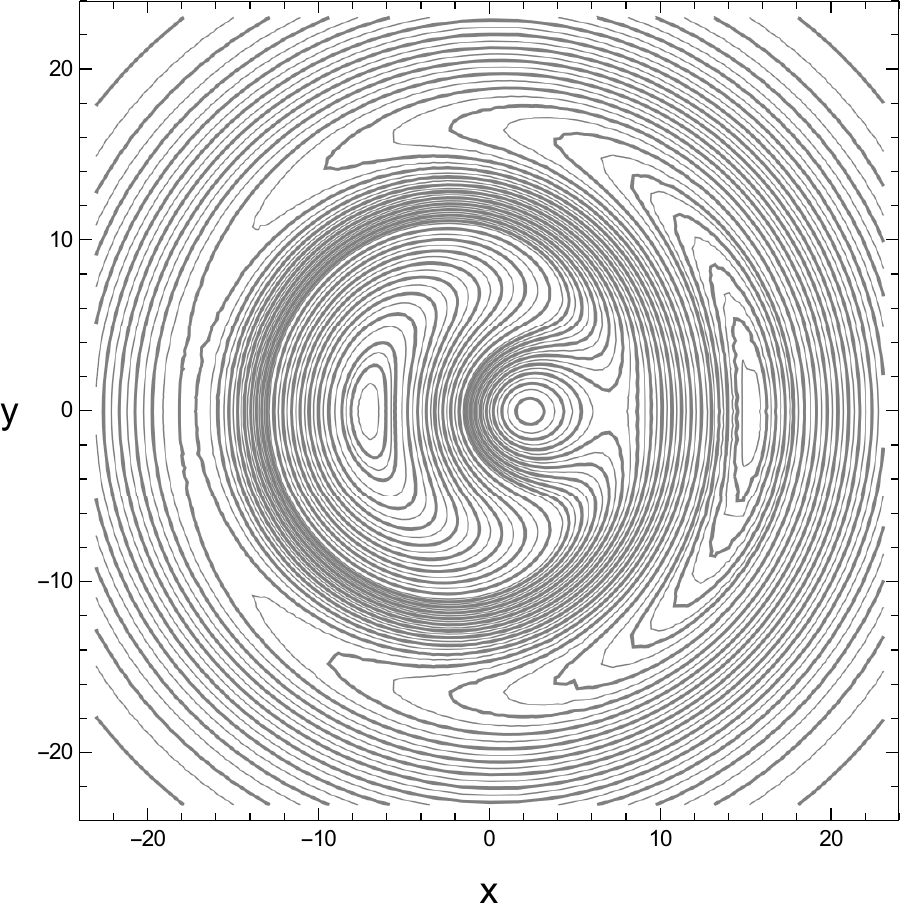}
 \caption{Poloidal cut of helicoidal magnetic surfaces $U(x ,y, z=0)=$const. for the helically symmetric equilibrium solution  (\ref{solh}). }
  \label{fig1}
 \end{figure}

\begin{figure}
 \centering
 \includegraphics[scale=0.9]{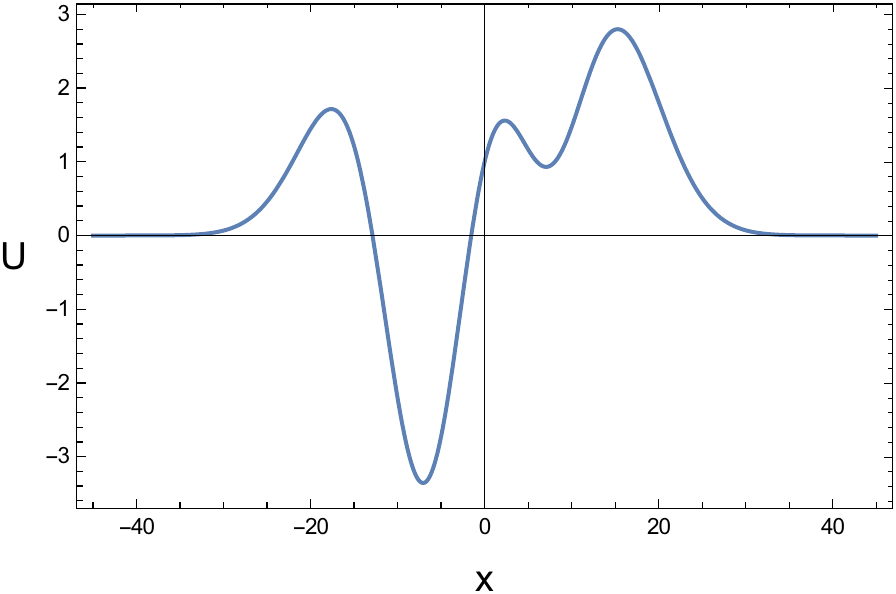}
 \caption{Profile of the function  $U(x ,y=0, z=0)=$const. for the helically symmetric equilibrium solution  (\ref{solh}). }
  \label{fig1a}
 \end{figure}

\begin{figure}
 \centering
 \includegraphics[scale=0.9]{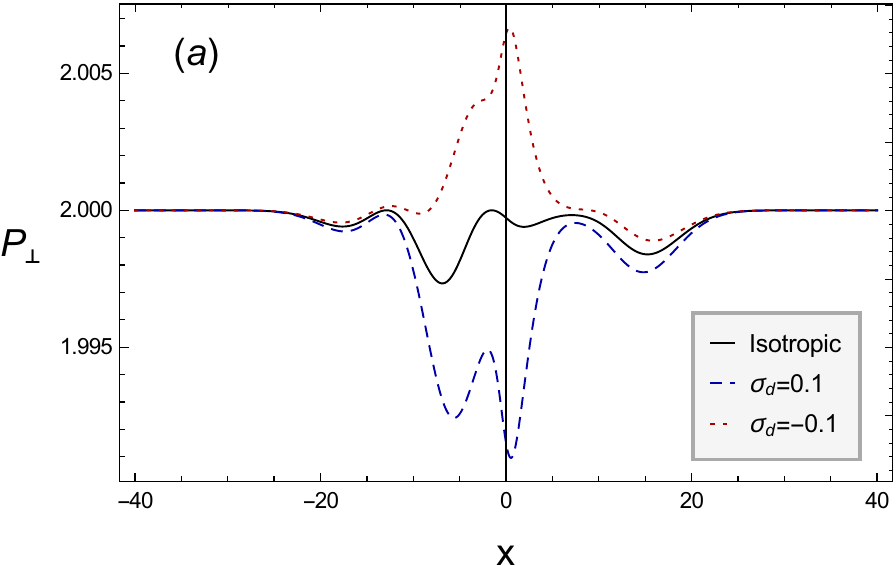}
 \includegraphics[scale=0.9]{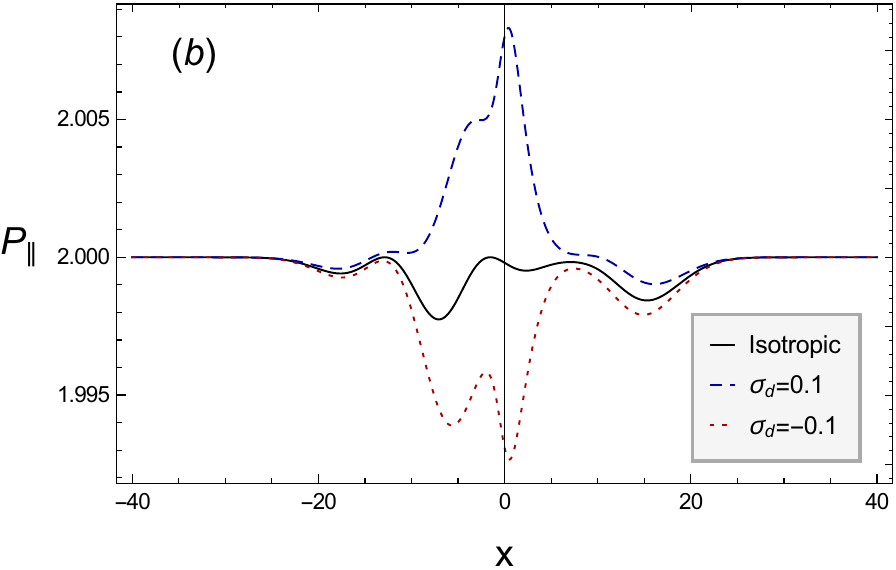}
 \caption{The profiles of the scalar pressures (a) $P_{\perp}(x,y=0,z=0)$ and (b) $P_{\parallel}(x,y=0,z=0)$ for the constructed stationary equilibria, for $M_{p_0}^2=10^{-3}$. The blue dashed curve correspond to $\sigma_d=0.1$, while the red dotted one to $\sigma _d=-0.1$. }
  \label{fig2}
 \end{figure}

\begin{figure}
 \centering
 \includegraphics[scale=0.9]{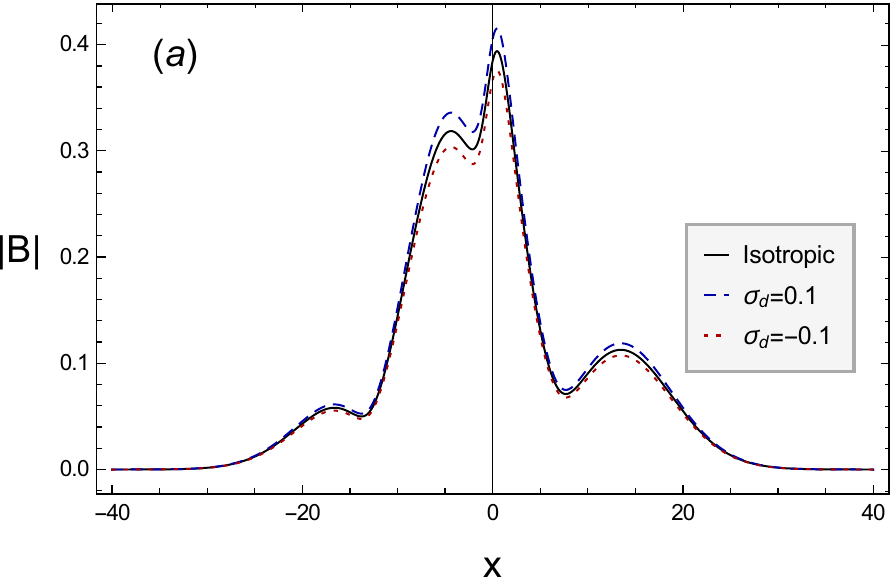}
 \includegraphics[scale=0.9]{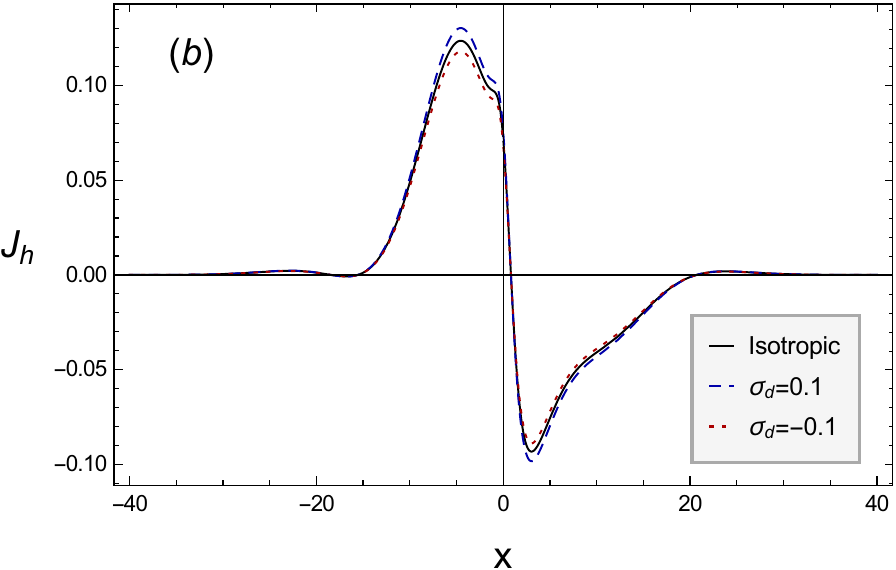}\\
\caption{Variation of (a) $|{\bf{B}}(x,y=0,z=0)|$ and (b) $J_{h}(x,y=0,z=0)$, for the  stationary equilibria constructed here, for $M_{p_0}^2=10^{-3}$ and the impact of pressure anisotropy on them for positive and negative values of $\sigma_d$. }
  \label{fig3}
 \end{figure}

\begin{figure}
 \centering
 \includegraphics[scale=0.9]{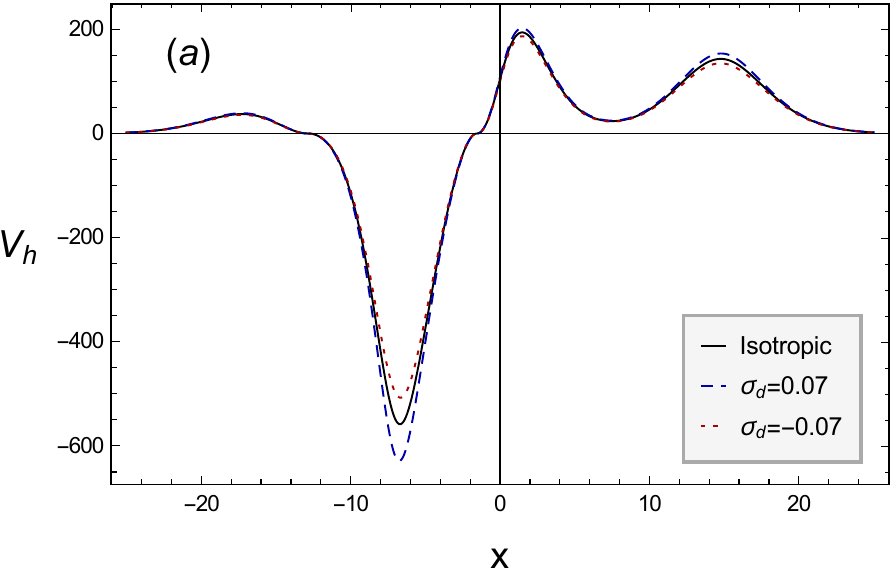}
 \includegraphics[scale=0.9]{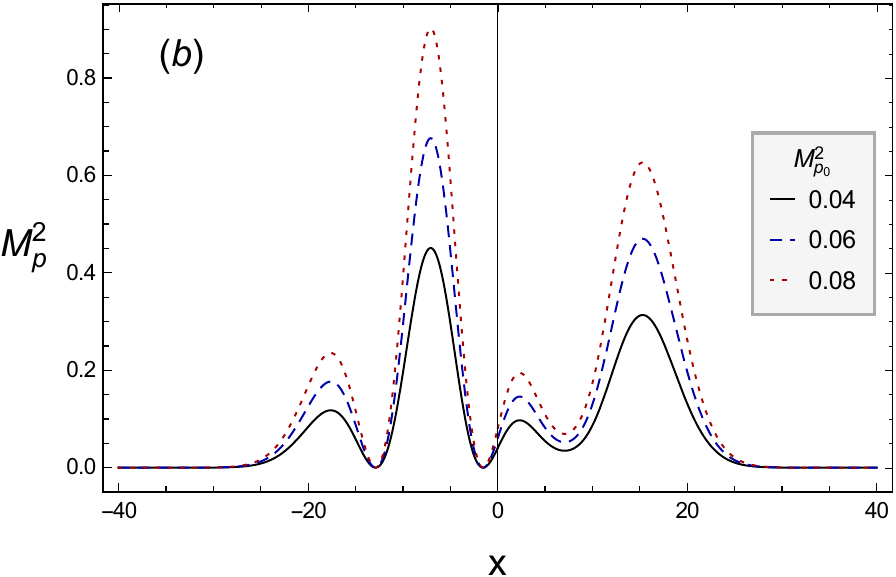}\\
\caption{Profiles of (a) the helicoidal velocity component for $M_{p_0}^2=0.06$, and the impact of pressure anisotropy through $\sigma _d$,  and  (b) the Mach function along the $x$-axis for different values of the flow parameter $M_{p_0}^2$. }
  \label{fig3a}
 \end{figure}

In order to construct specific equilibria, we restict our analysis to  solution (\ref{bogsol}) for $N=2, \, q=1, \, l=0$, and $w_1=7/(6\eta)$, which assumes the simpler form \citep[see][equation (4.2)]{bogo2}:
\begin{equation}
\label{solh}
U(r,u)=e^{-w_3 r^2}[1-10w_3r^2+8w_3^2r^4+c_{10}\cos(u/\eta)].
\end{equation}
Solution  (\ref{solh}) of the GS equation (\ref{GS}) associated with the relations  (\ref{helical}) and the profiles (\ref{ans})  defines a special class of exact helically symmetric equilibria  with incompressible flows with constant density and anisotropy functions, valid for any functional dependence of $M_p^2(U)$.  Therefore, to completely determine an equilibrium we employ the following profile for the Mach function
\begin{equation}
\label{mach}
M_p^2(U)=M_{p_0}^2U^2,
\end{equation}
 where $M_{p_0}$ is an arbitrary parameter. The magnetic surfaces of the above constructed equilibria on the cartesian plane $z=0$, for $w_3=0.01, \, c_{10}=0.5, \, w_2=2$, and $\eta =2.318$, are shown in figure \ref{fig1}; all dimensional quantities present in this section are measured in appropriate SI units. Also,  the profile of the flux function $U$ along the $x$-axis is given in figure \ref{fig1a}.
 In  figures \ref{fig1} and \ref{fig1a} it can be seen that  the plasma domain  consists of two sub-domains: an outer one consisting  of magnetic surfaces with circular poloidal cross-sections extending  up to infinity  
  ($r\rightarrow \infty$), and an  inner sub-domain  containing three lobes and a couple of saddle points (X-points). The inner $X$-point, corresponding  to a maximum of $U$ is located  at $x = -17.6$. Then on the right hand side of this first $X$-point are located successively  two lobes,   then the second $X$ point and farther outwards  the third lob. The respective magnetic axes of the lobes are located at $x = -7.07$, $x = 2.31$ and  $x = 15.3$ while the second $X$-point is located at $x = 7.07$.   Each helix composing such a helically symmetric configuration is characterized by a pitch length, equal to $2\pi \eta$, and a torsion  given by
 \begin{equation}
 \label{ht}
\tau= \frac{\eta}{r^2+\eta ^2}.
 \end{equation}
The above equilibrium can  model helically symmetric jets with anisotropic pressure, tending to become isotropic at very long distances   ($r\rightarrow \infty$). In more detail, for any $\sigma _d=\mbox{const.} \neq 0$ inside $\mathcal{D}$, the scalar pressures parallel and perpendicular to ${\bf{B}}$ are given by the relations 
 \begin{equation}
 \label{scalar}
 P_{\perp}=\mathcal{P}-\sigma _d\frac{{\bf{B}}^2}{2\mu _0}, \quad P_{\parallel}=\mathcal{P}+\sigma _d\frac{{\bf{B}}^2}{2\mu _0},
 \end{equation}
indicating  that when $\sigma _d >0$, its increase results in an  enhancement of $P_{\perp}$ while $P_{\parallel}$ decreases, and vice versa for $\sigma _d<0$, as can be seen at the profiles of  $P_{\parallel}$  and $P_\perp$ shown  in figure \ref{fig2}.
 In the limit of $r\rightarrow \infty$ the magnetic field,  current density and  velocity vanish and therefore the scalar pressures become equal each other, i.e.  $P_{\perp}=P_{\parallel}=w_2=\mbox{const.}$ because of  the second of (\ref{ans}) and (\ref{scalar}). Thus the configuration becomes in this limit isotropic. Note that this is compatible with a non zero value of $\sigma_d$ on account of the definition (\ref{sigma1}), which makes $\sigma_d$ indefinite in the limit of $r\rightarrow \infty$. Profiles of  the magnetic field magnitude, $B$, and the helicoidal component of the current density, $J_h$, are shown in figure \ref{fig3}. 
The values of $B$ becomes greater (lower) for $\sigma_d>0$ ($\sigma _d<0$), in connection with  a diamagnetic (paramagnetic) behavior. The helicoidal current density,  $J_h$,  reverses near the origin and  becomes more peaked (hollow) for $\sigma_d>0$ ($\sigma _d<0$). In addition profiles of the helicoidal velocity component, $V_h$ and the Mach function, $M_p^2$, are provided in figure \ref{fig3a}. 
  It is noted that $V_h$ reverses in the region of the left-lobe, where $U$ becomes negative, and that the  flow  in terms of the Mach function, $M_p^2$, strengthens the impact of pressure anisotropy  for $\sigma_d>0$, due to the factor $1-\sigma_d-M_p^2$ (cf. equation (\ref{GS})).  In this respect, it is expected that the increase of the parameter $M_{p_{0}}^2$ has the same impact on $V_h$ as that shown in figure \ref{fig3a}(a) for $\sigma_d >0$.
 
\begin{figure}
 \centering
 \includegraphics[scale=0.8]{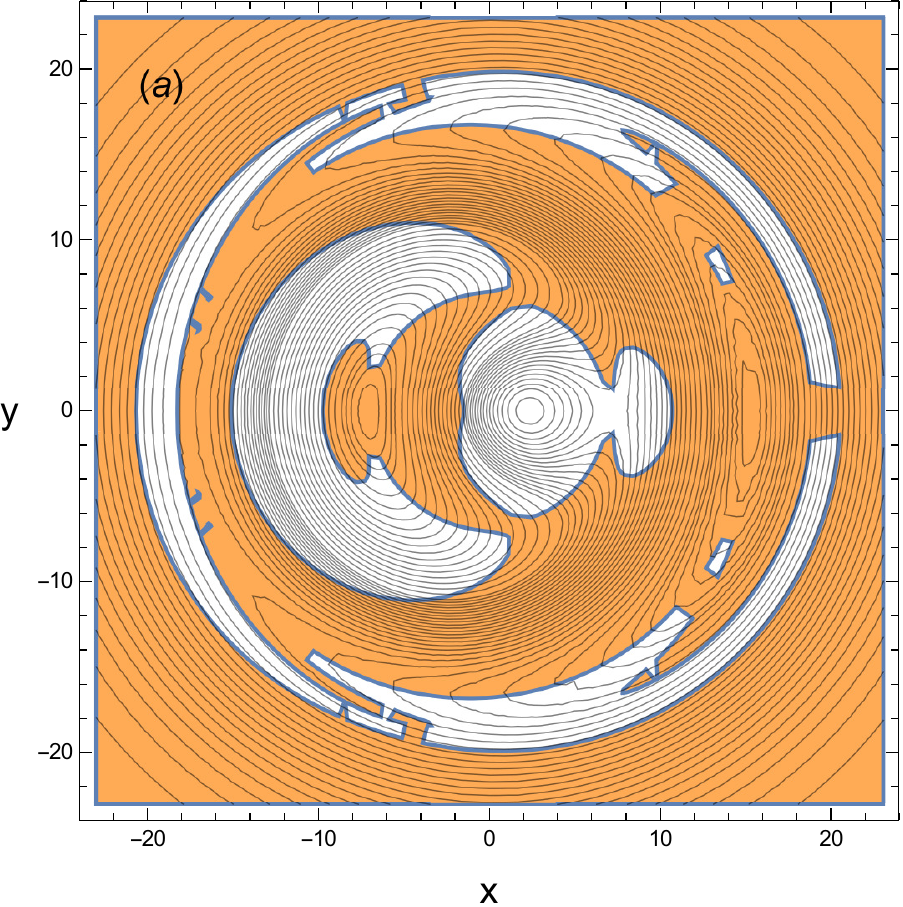}
 \includegraphics[scale=0.9]{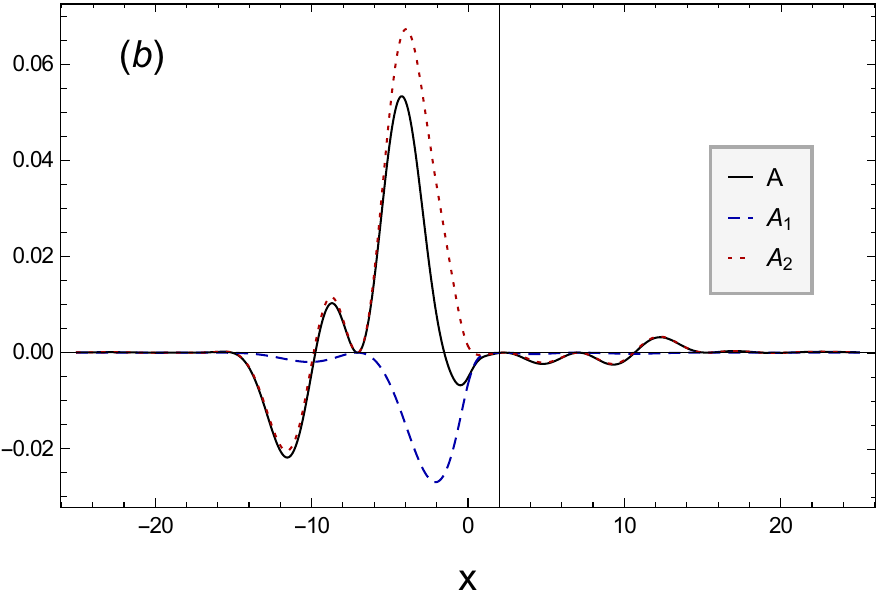}
 \caption{(a): For the static anisotropic helically symmetric equilibrium  ($\sigma_d=M_{p_0}^2=0$)  the stability condition $\mathcal{A}\geq 0$ is satisfied  in the orange coloured regions.  (b): The term $A_2$ has a stabilising effect (red-dotted curve)  which  counteracts  the destabilising one of  $A_1$ (blue-dashed curve),  so that the quantity $\mathcal{A}=A_1+A_2$ indicated by the black-straight curve becomes  positive in the aforementioned orange coloured regions. }
 \label{fig4}
 \end{figure}

In what follows, we employ the derived suffiecient condition by calculating the quantity $\mathcal{A}=A_1+A_2+A_3$ of equation (\ref{Au}) (in connection with equations (\ref{Alpha}), (\ref{A1})-(\ref{A3}) in the $\psi$-space) for the helically symmetric equilibria, defined by the relations (\ref{helical}), (\ref{ans}) and (\ref{solh})-(\ref{scalar}), in order to examine the impact of the pressure anisotropy, flow and  torsion on their linear stability, through the variation of the parameters $\sigma_d, \, M_{p_0}, \, \mbox{and}\, \eta$, respectively.
We recall  that the aforementioned relations were obtained by applying the integral transformation (\ref{Utrans}) \citep[see][]{eva2};  in this respect the condition (\ref{con1}) is trivially satisfied. Figure \ref{fig4}(a) shows that the condition $\mathcal{A}\geq 0$ in the absence of pressure anisotropy and flow, $\sigma_d=M_{p_0}^2=0$, is satisfied in a broad region including the outer domain and the two magnetic axes located on the  left and right side  of the  origin  $(x=y=0)$. In these regions, the term $A_2$ being  stabilising  surpasses  the destabilising term $A_1$ as shown in figure \ref{fig4}(b) (while in this case the flow term $A_3$ vanishes). However, the condition is satisfied neither near the central magnetic axis, where $J_h$ reverses and $U$ is negative, nor in a small region located at the left side of the magnetic axis of the left lobe, in which $V_h$ reverses; 
in this respect it should be  noted that since the  condition is only sufficient, the white colored regions in figure  \ref{fig4}(a) and in  the figures to follow, where $\mathcal{A}< 0$, do not necessarily imply instability. Thus we will consider only  regions in which the condition $\mathcal{A}\geq 0$ is satisfied. 

\begin{figure}
 \centering
 \includegraphics[scale=0.9]{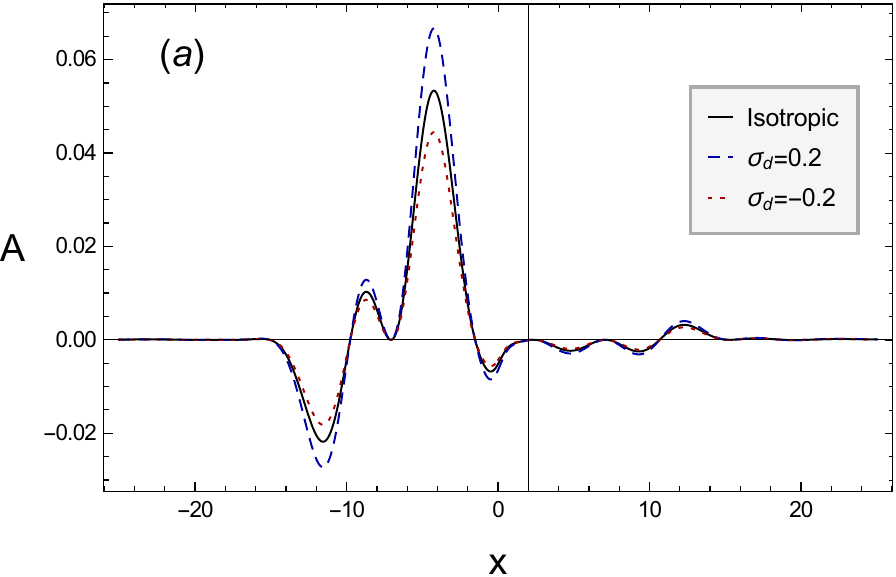}
 \includegraphics[scale=0.9]{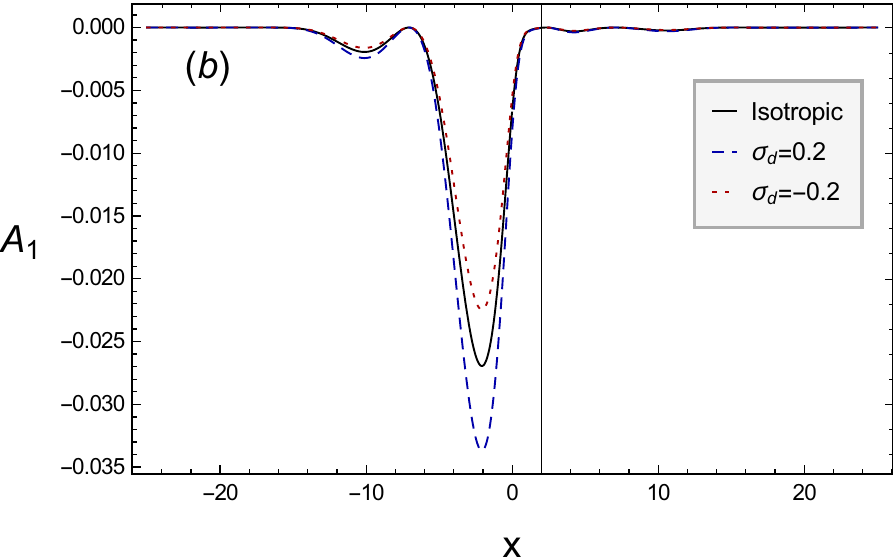}
 \includegraphics[scale=0.9]{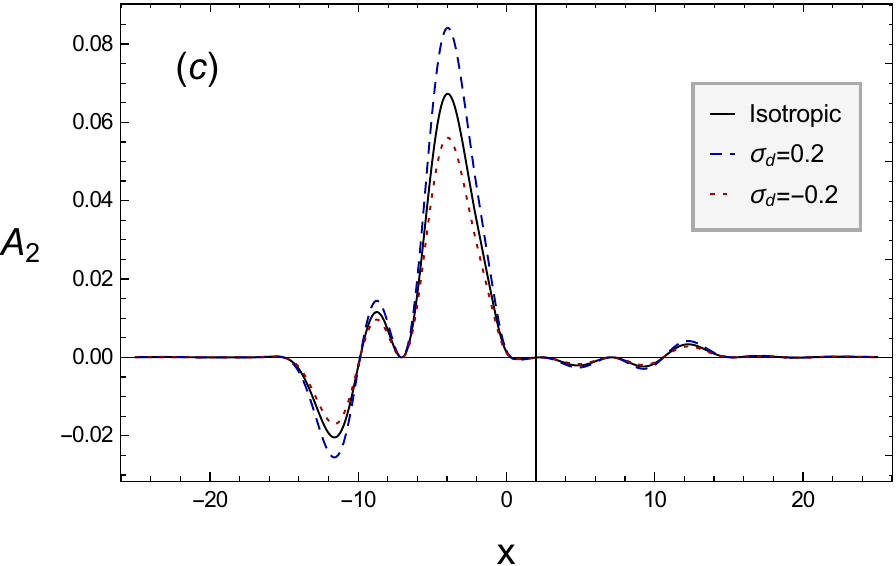}
 \caption{The impact of pressure anisotropy on the quantities (a) $\mathcal{A}$, (b) $A_1$  and (c) $A_2$  in the absence of flow  ($M_{p_0}=0$) for $\sigma_d=0.2$ (blue-dashed curves) and $\sigma_d=-0.2$  (red-dotted curve). For comparison  are also given the respective isotropic black continuous curves. In the regions where  $\mathcal{A}\geq 0$ this impact is stabilising for $\sigma_d>0$ and destabilising for $\sigma_d<0$. }
 \label{fig5}
 \end{figure}

\begin{figure}
 \centering
 \includegraphics[scale=0.8]{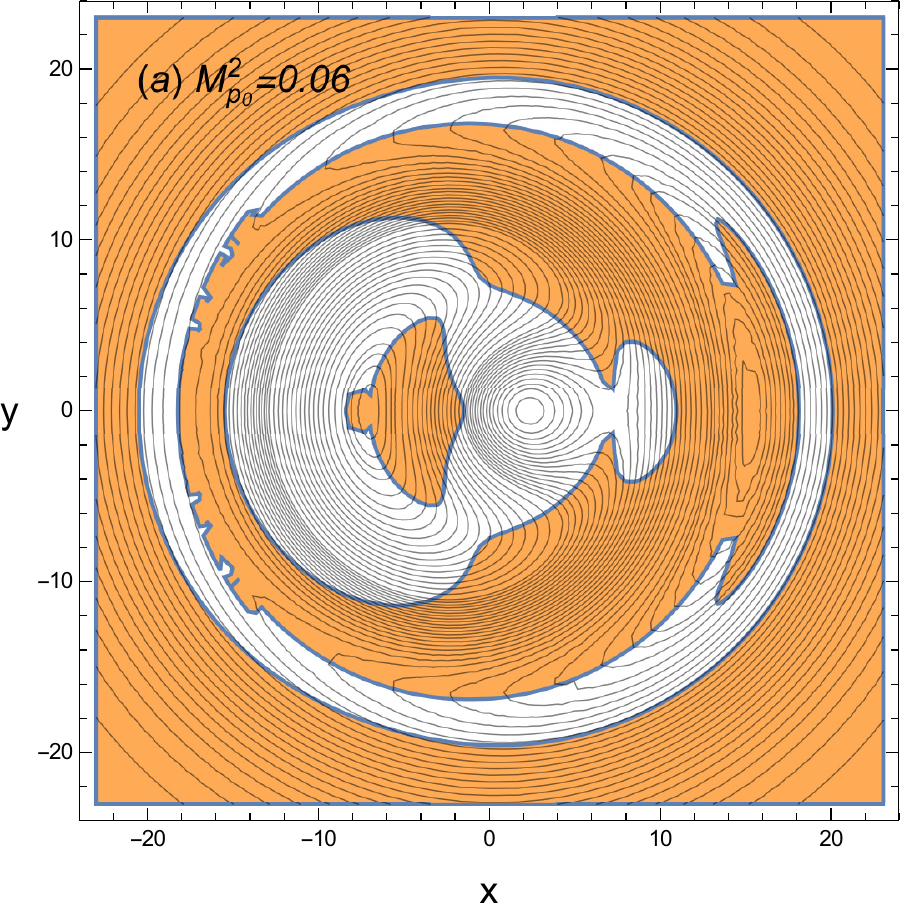}
 \includegraphics[scale=0.8]{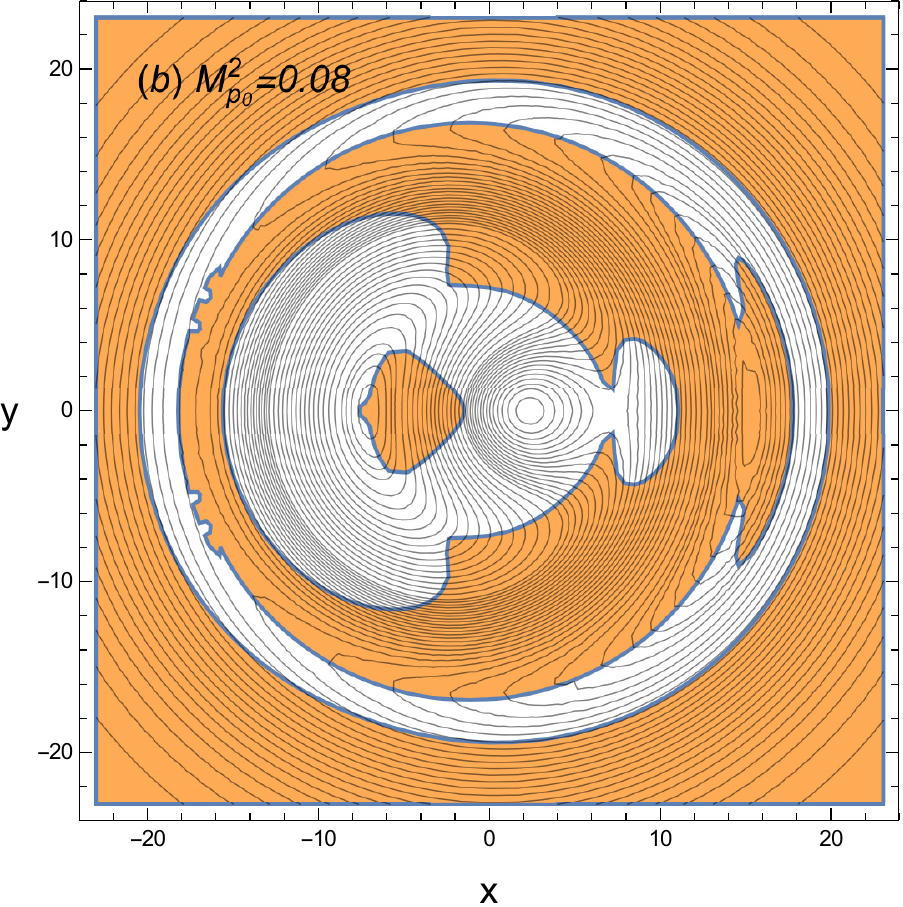}
 \includegraphics[scale=0.9]{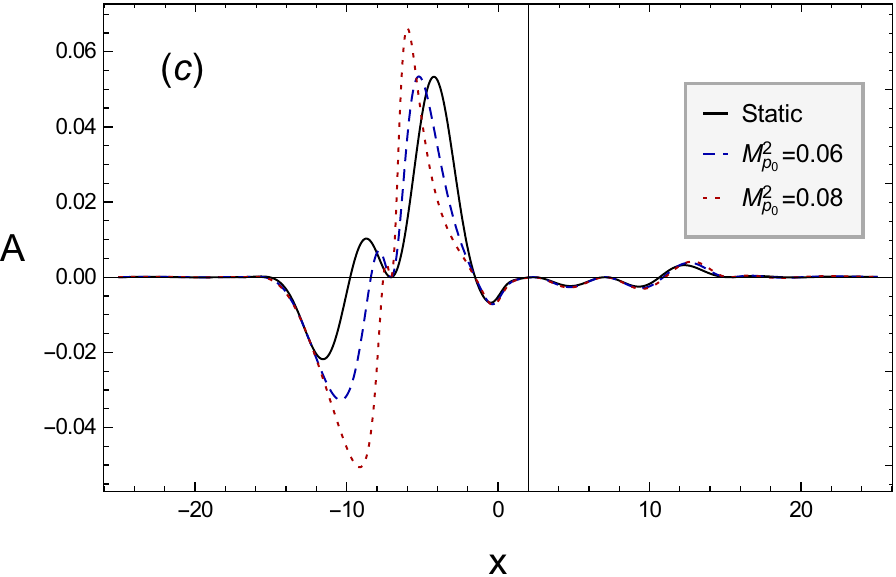}
 \caption{Impact of the flow through $M_{p_0}^2$ in the central orange colored region where the stability condition  $\mathcal{A}\geq 0$ is satisfied in comparison with  the respective static isotropic  equilibrium. The maximum used value of the parameter $M_{p_0}$ for which all the pressures remain positive is 0.08.}
 \label{fig6}
 \end{figure}

\begin{figure}
 \centering
 \includegraphics[scale=0.9]{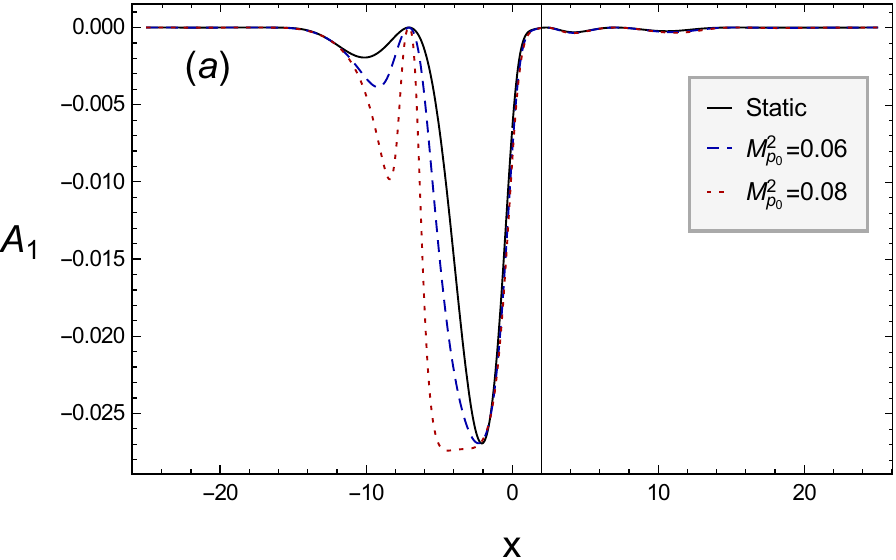}
 \includegraphics[scale=0.9]{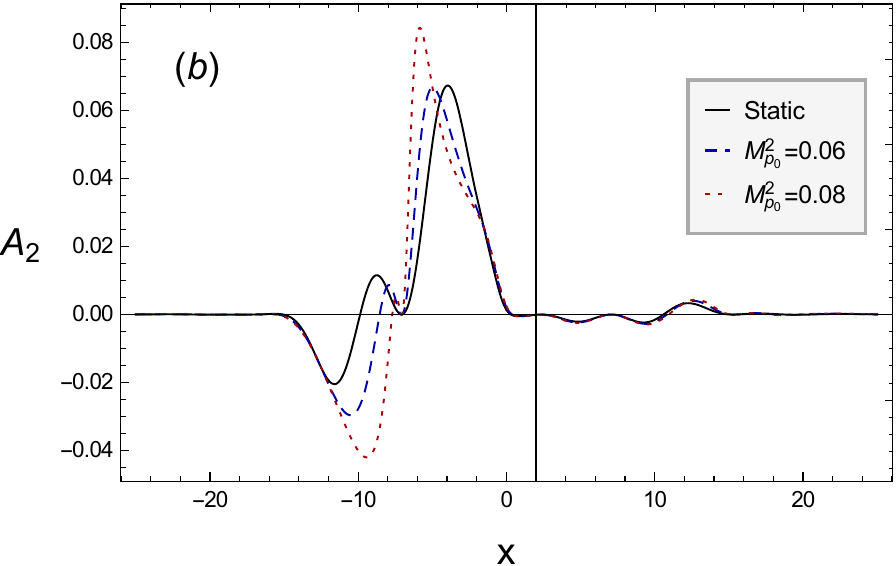}
 \includegraphics[scale=0.9]{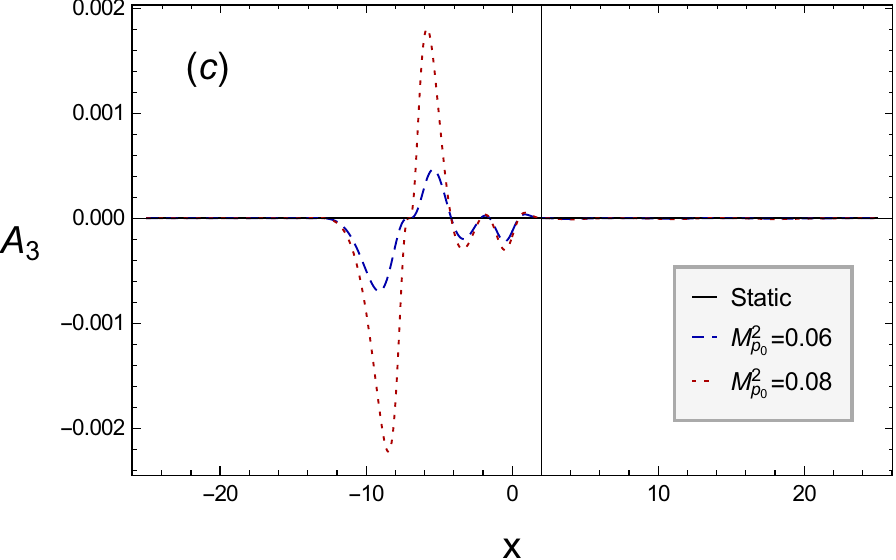}
 \caption{The impact of the flow parameter $M_{p_0}^2$ on the terms: (a) $A_1$, (b) $A_2$, and (c) $A_3$, for $\sigma_d=0$.}
 \label{fig7}
 \end{figure}

\begin{figure}
 \centering
 \includegraphics[scale=0.9]{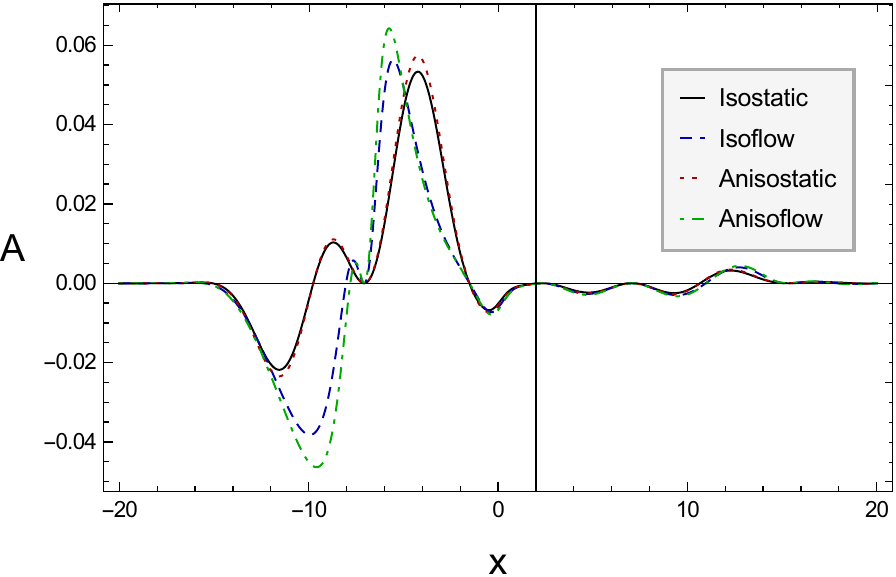}
 \caption{The overall stabilising  impact of pressure anisotropy in combination with flow on the stability of the constructed helically symmetric equilibria. The pertinent  parametric values employed are as follows: (straight black curve) $\sigma_d=M_{p_0}^2=0$, (blue dashed curve) $\sigma_d=0, \, M_{p_0}^2=0.07$, (red dotted curve) $\sigma_d=0.07, \, M_{p_0}^2=0$, and (green dot-dashed curve) $\sigma_d=M_{p_0}^2=0.07$.}
 \label{fig8}
 \end{figure}

The presence of pressure anisotropy  does not affect the isotropic stability map of  figure \ref{fig4}(a), as it is clearly indicated in the profiles of figure \ref{fig5}. However, it affects the values of A. Specifically, in the regions where $\mathcal{A}\geq 0$,   for $P_{\parallel}>P_{\perp}$ ($\sigma _d>0$)  the anisotropy has a stabilising impact, in the sense that the maximum values of $\mathcal{A}$ become larger than the respective isotropic ones, and a destabilising effect  for $\sigma_d<0$.  
These characteristics  are illustrated   in figure \ref{fig5}.
In addition, it is found that the  flow in terms of $M_p^2$ has a peculiar effect on stability. Specifically, on the one hand it results in  shrinking  the orange colored  area located on the left hand side of the first lobe, where the helicoidal velocity reverses, as it can be seen in figure \ref{fig6}(a),(b). This  shrinking is connected with a destabilising effect of  $M_{p0}^2$ on  both terms $A_1$ and $A_2$ as shown in figure \ref{fig7}(a),(b). On the other hand the flow has a stabilising effect, similar to that of $\sigma_d>0$,  because the respective maximum values of   $\mathcal{A}$ get larger in this area as $M_p^2$ increases, as it can be seen in figure \ref{fig6}(c).
Also, the flow gives rise to a stabilising  contribution via  the term $A_3$.  However, this contribution in the region of interest is an order of magnitude lower than the destabilising impact of  $M_p^2$ on $A_1$ and $A_2$,  as can be seen in figure \ref{fig7}(c).
 Because of the stronger impact of the pressure anisotropy on  $\mathcal{A}$
 than the destabilising effect of the flow, the presence of both anisotropy and flow has an overall stabilising effect in terms of the region where the condition $\mathcal{A}\geq 0$ is satisfied and the maximum values of $\mathcal{A}$. This is shown in  figure \ref{fig8}. 
 
\begin{figure}
 \centering
 \includegraphics[scale=0.75]{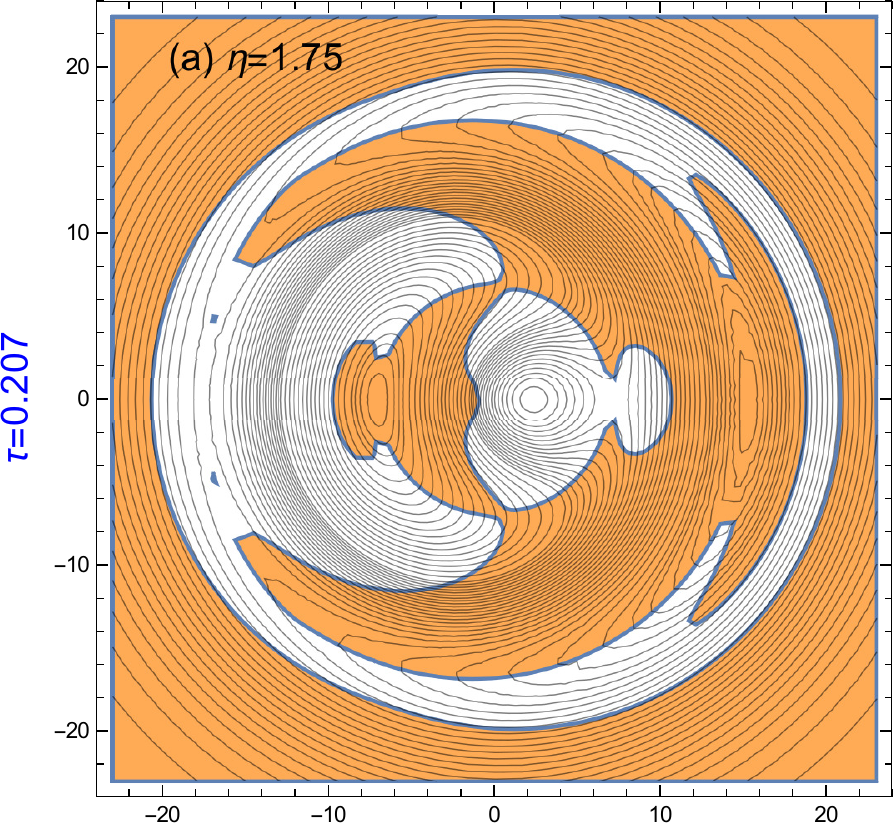}
 \includegraphics[scale=0.75]{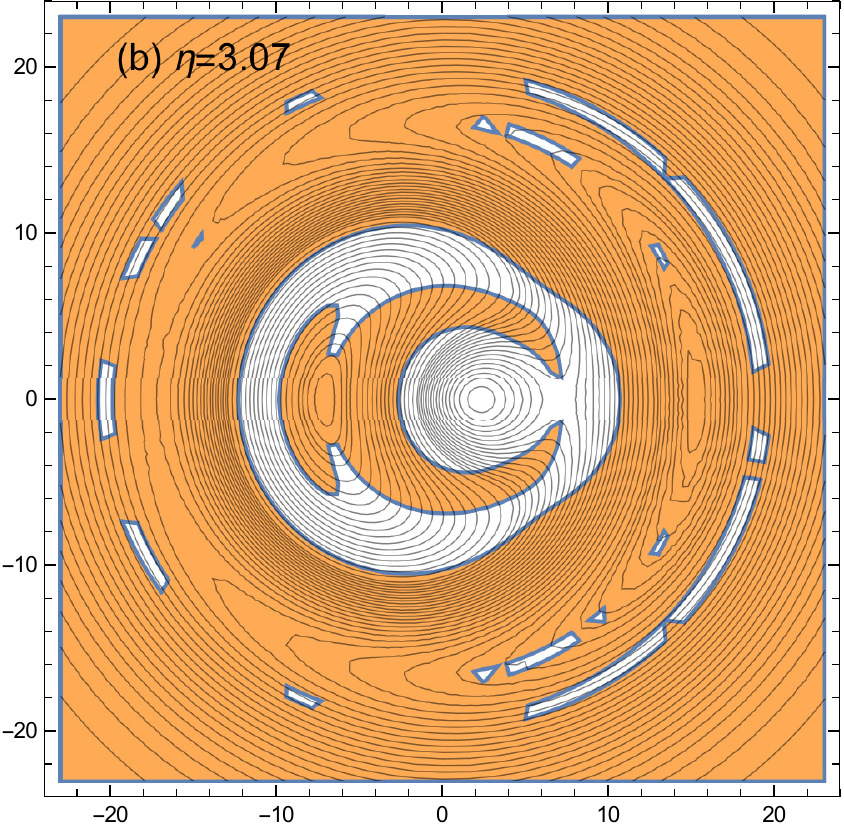}
 \includegraphics[scale=0.75]{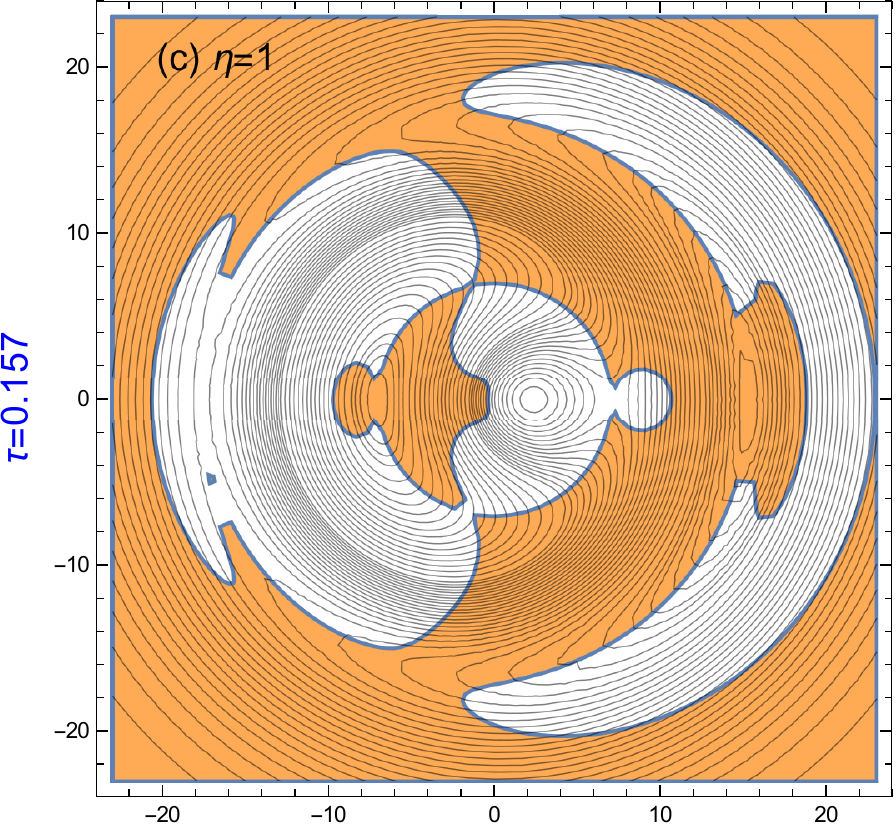}
 \includegraphics[scale=0.75]{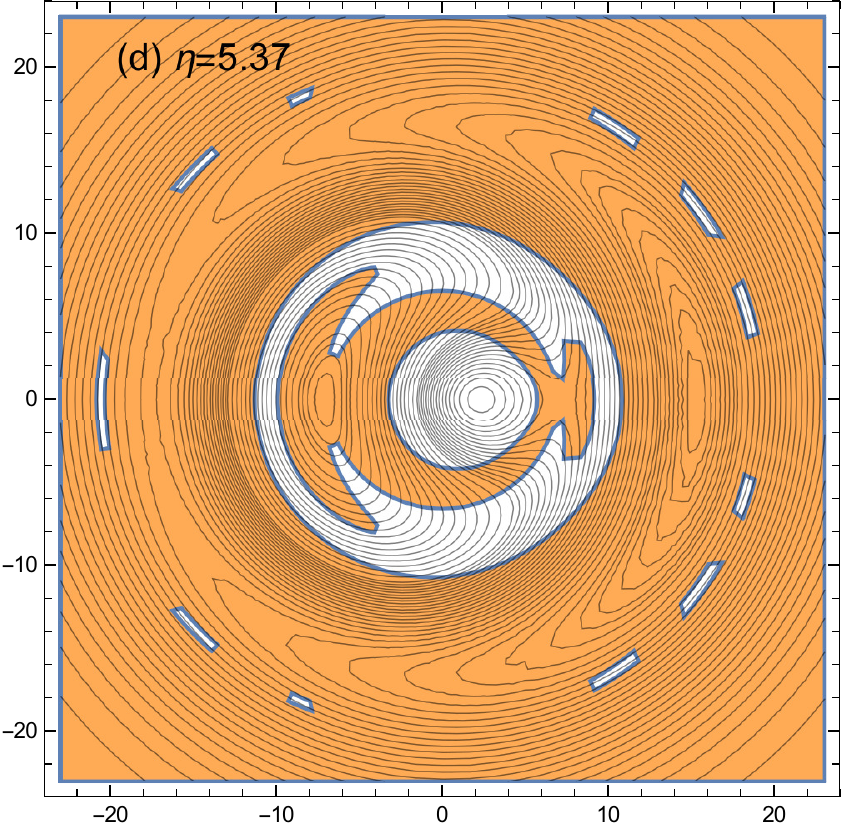}
 \includegraphics[scale=0.75]{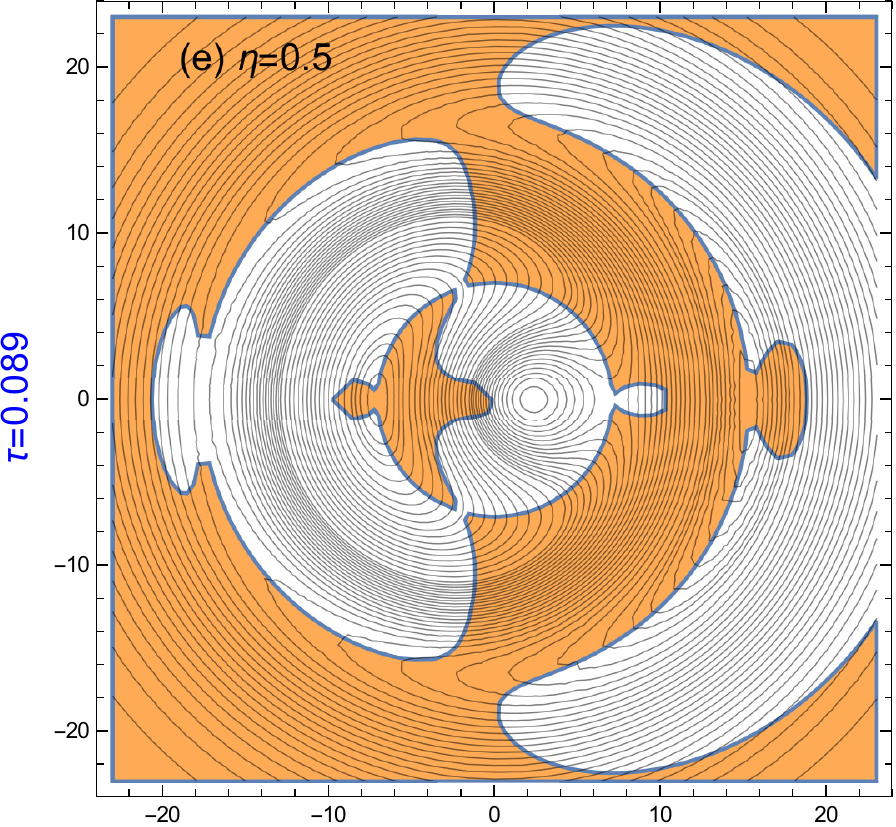}
 \includegraphics[scale=0.75]{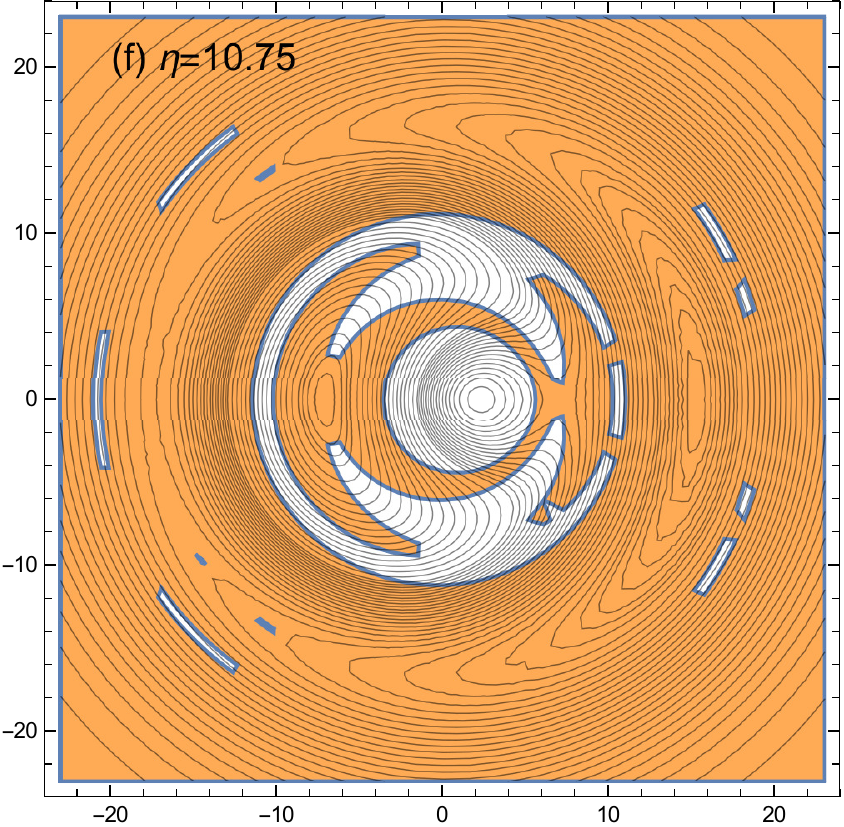}
 \caption{The impact of the torsion of the magnetic axis and of the picth on the linear stability of helically symmetric equilibria. Each pair of figures illustrate configurations that have same torsion, but different pitch values as explained in the text. }
 \label{fig9}
 \end{figure}

Finally, as concerns the impact of the torsion, $\tau$,  on the quantity $\mathcal{A}$, for a specific helix, defined by the equations $r=r_c=\mbox{const.}, \, u=u_c=\mbox{const.}$ (in helical coordinates), equation (\ref{ht}) implies that  $\tau$ depends only on the parameter $\eta$, which characterises  the pitch of that  helix. Inspection of equation (\ref{ht}) implies  that $\tau$ has an extremum for $\eta=r_c$, corresponding to the maximum torsion, $\tau _{max}=1/2r_c$. For example, for  the static, isotropic  equilibrium of  figure \ref{fig1} the helical magnetic axis of  the central  lobe  intersects the plane  $z=0$ at the position  $x_c=2.318, \, y_c=0$,  corresponding to  $r_c=2.318$, and has maximum torsion, $\tau _{max}=0.2157$. Therefore,  there can exist two different helically symmetric configurations with the same torsion but different pitches one for $\eta <r_c$ and the other $\eta >r_c$. However, the same torsion does not imply that these configurations have necessarily the same stability properties.
The regions in which   the condition $\mathcal{A}\geq 0$ is satisfied for the equilibrium  of figure \ref{fig1} is shown in figure \ref{fig4}.   We found the respective stability maps for   three pairs of equilibria, shown in figure \ref{fig9}. Each pair corresponds to the same torsion, $\tau < \tau _{max}$ but different pitch lengths, $\eta <r_c$ and $\eta >r_c$.
Specifically,  the  torsion and pitch values we employed in connection with these  equilibrium pairs are the following: (upper pair consisting of the figures \ref{fig9}(a),(b)) $\tau _{(a),(b)}=0.207, \, \eta _{(a)}=1.75, \, \eta _{(b)}=3.07$, (middle pair consisting of the figures \ref{fig9}(c),(d)) $\tau _{(c),(d)}=0.157, \, \eta _{(c)}=1, \, \eta _{(d)}=5.37$, and (lower pair consisting of the figures \ref{fig9}(e),(f)) $\tau _{(e),(f)}=0.089, \, \eta _{(e)}=0.5, \, \eta _{(f)}=10.75$.    The stability maps indicate that the condition  $\mathcal{A}\geq 0$ is satisfied in a wider region as the torsion decreases from its maximum value and for a given torsion $\mathcal{A}$  it gets larger as the  pitch length $\eta >r_c$ increases. Thus, we conclude that helical configurations with smaller torsion and bigger pitch lengths may have improved stability characteristics. This result is reasonable if one  considers the  limit  of zero torsion and infinite pitch length in which a helically symmetric plasma column degenerates into an one dimensional, cylindrical $\theta$-pinch. It is well known that such a configuration has favorable stability properties, since the safety factor approaches infinity. However, confinement in a $\theta$-pinch is not possible in the presence of toroidicity because the axial magnetic field becomes  purely toroidal. 

\section{Conclusions}\label{7}

We have derived a sufficient condition for  the linear stability of plasma equilibria for field-aligned incompressible flows in connection to plasmas of constant density and   pressure anisotropy such that the pressure difference $P^{*}_\parallel-P^{*}_\perp$ be proportional to the magnetic pressure,  by employing an  Energy Principle. Specifically, we have shown that the  linear stability of such kind of equilibria,  being guaranteed when the  functional of the perturbation potential energy, relates to the   sign of a function $\mathcal{A}$ [equation (\ref{A})] which depends only on equilibrium quantities. According to that  condition,  any equilibrium   is linearly stable to small three-dimensional perturbations whenever (i)  the sum of the anisotropy function plus the Mach function of the  equilibrium velocity colinear with  the magnetic field is not larger than  unity [equation (\ref{con1})] and (ii) $\mathcal{A}$ is non-negative [equation (\ref{con2})]. This condition  generalises  the sufficient condition derived  in \cite{throumtaso} for respective equilibria with isotropic pressure.

The aforementioned condition can be applied to any plasma steady state without  geometrical restriction, that is, it can be employed for three-dimensional equilibria. The quantity $\mathcal{A}$ involved  consists of three  physically interpretable terms. The first of these terms,  being  always negative and  therefore  destabilising, may relate to current driven instabilities. The other two terms may have either positive or negative contribution to $\mathcal{A}$ depending on the   characteristics of the background equilibria. The second term relates to the magnetic shear, while the third term relates to the velocity shear and to the variation of the total pressure perpendicular to the magnetic surfaces; the latter term  vanishes for static equilibria. 

In addition, we have shown that if a given equilibrium with field-aligned incompressible flows of constant mass density and constant  pressure anisotropy function fulfills  the aforementioned  condition, and therefore is  linearly stable,  then all the  families  of  equilibria obtained  by  the application of the symmetry transformations presented in \cite{eva1} on the original equilibrium, are also linearly stable, provided that  a parameter, $C$, appearing in these transformations, is positive definite.

At last, we applied the aforementioned sufficient condition to a special class of helically symmetric equilibrium solutions  describing astrophysical jets  in order to examine the impact of  flow,  pressure anisotropy as well as  of the torsion and  pitch of certain equilibrium-configuration helices   on  stability. For this class of equilibria we have found that both the flow and the anisotropy can have noticeable impact   on  stability, which  in different plasma regions can be  either stabilising or destabilising; the impact of pressure anisotropy is stronger that that of the flow. Specifically, in the regions where the stability condition $\mathcal{A}\geq0$ is satisfied the combined effect of flow and anisotropy is stabilising when  $P_{\parallel}>P_\perp$.  Finally, the results indicate that helically symmetric equilibrium configurations with smaller torsion and larger  pitch length  are favored in terms of stability.

It is interesting to pursue potential extensions of the sufficient condition derived here to steady states with compressible flows or/and more physically relevant pressure anisotropy. This requires replacing incompressibilty and the assumption of constant pressure anisotropy function by alternative equations of state on the understanding that finding self consistently more appropriate equations of state (or energy equations) associated  with the pressure tensor elements  $P_\parallel$ and $P_\perp$ relates to the  tough  closure problem of a hybrid kinetic-fluid model, e.g. \cite{snha1997,ra2015a}. 

\section*{Acknowledgements}

This study was performed within the framework of the EUROfusion Consortium and has received funding from the Euratom research and training programme 2014-2018 and 2019-2020
under grant agreement No 633053 as well as  the National Program for the Controlled Thermonuclear Fusion, Hellenic Republic. The views and opinions expressed herein do not necessarily reflect those of the European Commission. A.E. has in part
been financially supported by the General Secretariat for Research and
Technology (GSRT) and the Hellenic Foundation for Research and
Innovation (HFRI).

\end{document}